\documentclass[a4paper,fleqn,usenatbib]{mnras}  
\usepackage[T1]{fontenc}
\usepackage{ae,aecompl}
\usepackage{graphicx}
\usepackage{amsmath}    
\usepackage{amssymb}    
\usepackage{epstopdf}
\bibpunct{(}{)}{;}{a}{}{,}
\newcommand{\cep} {CoRoT~0102618121}
\newcommand{\cepdue} {V376~Mon}
\newcommand{\fu} {$f_1$}
\newcommand{\cd} {d$^{-1}$}
\newcommand{\bc} {$b_C$}
\newcommand{\rc} {$r_C$}
\newcommand{\gc} {$g_C$}
\newcommand{\rdu} {$R_{21}$}
\newcommand{\fdu} {$\phi_{21}$}
\newcommand{\ftu} {$\phi_{31}$}
\newcommand{\fqu} {$\phi_{41}$}
\newcommand{\tmax} {T$_{\rm max}$}
\newcommand{\kms} {km\,s$^{-1}$}
\newcommand{\teff} {$T_\mathrm{eff}$}

   \title[The seven CoRoT Cepheids]{CoRoT space photometry of seven Cepheids\thanks{Partially based on observations collected
at La Silla Observatory, ESO (Chile) with the HARPS spectrograph at the 3.6-m telescope under
programme LP185.D-0056}}

   \author[E. Poretti et al.]{E.~Poretti,$^{1,2,3}$\thanks{E-mail: ennio.poretti@brera.inaf.it} 
J.~F. Le Borgne,$^{2,3}$
M.~Rainer,$^{1}$
A.~Baglin,$^{4}$
J.~M.~Benk\H{o},$^{5}$
\newauthor 
J.~Debosscher,$^{6}$
and W.W.~Weiss$^{7}$
\\
   $^{1}${INAF-Osservatorio Astronomico di Brera, Via E. Bianchi 46, I-23807 Merate, Italy}\\
$^{2}$Universit\'e de Toulouse; UPS-OMP; IRAP;  F-31400 Toulouse, France\\
$^{3}$CNRS; IRAP; 14, avenue Edouard Belin, F-31400 Toulouse, France\\
$^{4}$LESIA, Universit\'e Pierre et Marie Curie, Universit\'e Denis Diderot, Observatoire de
     Paris, F-92195 Meudon Cedex, France\\
$^{5}$Konkoly Observatory, MTA CSFK, Konkoly Thege u. 15-17., H-1121 Budapest, Hungary\\
$^{6}$Instituut Voor Sterrenkunde, Catholic University of Leuven, Celestijnenlaan 200D, B-3001
     Leuven, Belgium\\
$^{7}$ Institute of Astronomy, University of Vienna, T\"urkenschanzstrasse 17, A-1180 Vienna, Austria\\
             }

   \date{Received ...; accepted ...}
\pubyear{2015}

\begin{document}
\label{firstpage}
\pagerange{\pageref{firstpage}--\pageref{lastpage}}
\maketitle

  \begin{abstract}
A few Galactic classical Cepheids were observed in the programmes of 
space missions as {\it Coriolis}, {\it MOST} and {\it Kepler}. An appealing opportunity  was to detect
additional nonradial modes, thus opening
the possibility to perform asteroseismic studies and making the pulsational
content of Galactic Cepheids more similar to that of Magellanic Clouds ones.
However, only hints of cycle-to-cycle variations were found, without any strict periodicity.
In this context the potential of the CoRoT exoplanetary data base was not fully exploited
despite the wide area covered on the Galactic plane. Therefore, we investigated all the candidate Cepheids
pointed out by the automatic classification of the CoRoT curves.
At the end we could identify seven bona-fide Cepheids.
   The light curves were investigated to remove some instrumental effects.
The frequency analysis was particularly delicate since these small effects can
be enhanced by the  large amplitude, resulting in the presence of significant,
but spurious, peaks in the power spectrum.
Indeed,  the careful evaluation of a very attracting peak
in the spectra of \cep\, allowed us to certify its spurious 
origin.
   Once that the instrumental effects were properly removed, no additional mode was detected.
On the other hand,
cycle-to-cycle variations of the Fourier parameters were observed, but very small and always within $\pm3\sigma$. 
Among the seven Cepheids, there are two 
 Pop.~I first-overtone pulsators, four 
Pop.~I fundamental mode pulsators, and one 
Pop.~II star. 
The CoRoT colours allowed us to
measure that times of maximum brightness occur a little earlier (about 0.01~period) at short wavelengths than at long ones.
  \end{abstract}
\section{Introduction}

The use of Cepheids as stellar candles requires a deep knowledge of
their pulsational characteristics. The separation
between fundamental ($F$-mode)  and first-overtone ($1O$-mode)
radial pulsators \citep{apr} is still the  most relevant
distinction to be taken into account to calibrate the period-luminosity
relations.  About the regularity
of their variations, ground-based photometry of V473~Lyr \citep{burki}
and Polaris  \citep{evans} showed continuous changes in amplitude
and phase.  \citet{molnar} established that V473~Lyr 
pulsates in the second radial overtone and shows strong amplitude and phase
modulations on two long periods (1205$\pm$3 and 5300$\pm$150~d).
The slow changes 
in the light curve of Polaris on a time baseline of 4.5~yr were precisely recorded
by space photometry using the {\it Coriolis} spacecraft. \citet{smei}
connected them to a quite fast evolutionary phase. However, \citet{bruntt}
proposed a different interpretation pointing out how
the amplitude change is cyclic rather than monotonic and also suggesting
an excess of power in the 2-6~d range due to the granulation.

The situation was expected to change with the advent of high-precision space
mission. 
A detailed analysis of the {\it MOST} observations of SZ~Tau and RT~Aur revealed
variations from cycle to cycle as a function of the pulsation phase \citep{most}.
The stars were both monitored for less than seven consecutive cycles:
the $1O$-Cepheid  SZ Tau showed a pulsation less stable
than the $F$-Cepheid RT Aur. The historical analysis of the times
of maximum brightness proved that  both stars experienced 
period variations  on a time-scale of decades and that those of the overtone 
mode resulted to be more erratic than the $F-mode$ ones. 

In the case of {\it Kepler}, there was only one Cepheid in the
field-of-view, i.e., V1154~Cyg. \citet{v1154} provided  a very detailed 
analysis of the data acquired in the first 600~d (i.e., about
120 cycles) of the mission, detecting significant cycle-to-cycle fluctuations.
A very slight correlation was also found between the Fourier parameters and
the O-C values, i.e., the difference between the {\it observed}
and {\it calculated} times of maximum brightness (\tmax).
This suggested that the O-C variations (up to 30~min) might be due to 
instabilities in the light curve shape. On the basis of these observational
results, \citet{neilson} suggested that convection and hot spots
can explain the observed period jitter, thus supporting the \citet{bruntt}
interpretation  of  the Polaris changes. The clues about the changes 
in the light curve of V1154~Cyg were strengthened by the analysis of new
{\it Kepler} measurements (up to Q17 quarter): 
the analysis of the Fourier parameters supplied a robust determination
of a periodicity of 158.6~d \citep*{poster}.

About ground-based observations, \citet{anderson} discovered significant modulations in the radial velocity
curves of some Cepheids. These modulations appear to be different from those
reported on V1154~Cyg, since they manifest as shape and amplitude variations
on time-scales of years for short-period Cepheids (like V1154 Cyg) and cycle-to-cycle variations
in the long-period Cepheids. 

A common result among all these investigations is that no 
nonradial mode was identified in the light or radial velocity curves 
of Galactic Cepheids. Hence the 
counterparts of the subclasses detected in the Large Magellanic Cloud   
\citep[LMC; for a review see ][]{pavel} are still missing.  
However, an important player has not yet contributed, i.e., 
the CoRoT data base of the light curves collected in the exoplanetary fields.  
Such data base 
already  proved to be  a real treasure allowing the discovery of  
a unique triple-mode Cepheid in the outskirts of the Milky Way \citep*{pbw}.
In this paper we describe the new elements brought by the CoRoT photometry 
to the hot topic of the stability of the Cepheid light curves.

\section{CoRoT data and spectroscopic follow-up}

The first new Cepheid discovered by CoRoT was
2MASS J06415168-0120059$\equiv$\cep, a 12$^{\rm th}$-mag star located at
$\alpha$ = $06^{\mathrm h}$\,$41^{\mathrm m}$\,$51\fs{685}$,
$\delta$ = --01\degr\,20\arcmin\,05\farcs{960} (J2000). It
was observed during the first Long Run in the Anticentre
direction (LRa01)
and then in the LRa06 four years later (Table~\ref{runs}).
The classification of the type of variability  was a little troublesome.
The star was assigned to the RR Lyr Working Group, but  
the preliminary analysis immediately pointed out that the star was not a RR~Lyr variable
since the period was too long. The Cepheid variability was much more probable, but some
doubts still remained, based on the classification as M1V type 
reported in the header of the FITS file of the LRa01 data. Indeed, the
``CoRoT Variability Classifier" automated supervised method \citep[CVC; ][]{cvc}
suggested an active, rotational variable as first hypothesis and 
a classical Cepheid as a second one. The spectral classification of \cep\, was
revised as A5V in the current version of the  Exo-Dat \citep{exodat} catalogue
and consequently reported  in the header of the more recent LRa06 data.

While the analysis of \cep\, was in progress, we proposed known Cepheids as CoRoT
targets by answering the Announcements of Opportunity regularly issued before any run.
In such a way V376 Mon$\equiv$CoRoT~0300003751 and DP~Mon$\equiv$CoRoT 0221644967  
were monitored  in the LRa02 and SRa02 runs, respectively (Table~\ref{runs}). 
However, it should be  expected that other Cepheids would be discovered thanks to
the deep survey performed by CoRoT on the Galactic plane. 

To search for these new Cepheids, we first scrutinized 
all the new variables discovered with the 
Berlin Exoplanet Search Telescope (BEST) that have been classified as candidate Cepheids \citep{bestI, bestII,
bestIII,bestIV,bestV,bestVI} and actually observed with CoRoT. We noticed that the variability of V376~Mon 
was recorded \citep{bestV} with the correct period, i.e., 1.652~d \citep{berdni}, not the
0.623~d reported in the discovery paper \citep{arno}. Unfortunately none of other BEST candidates
were confirmed as a new Cepheid, since they turned out to be rotational variables.
The misclassification  is probably due to the too relaxed criterion used to classify
a new variable as Cepheid, i.e., a periodic variability with an amplitude larger than 0.10~mag.
Indeed, such an amplitude threshold can be reached by active spotted stars. 

As a second step in the data mining of the CoRoT data base, we selected all the 
variables for which the CVC method proposed a classification 
as Cepheid, independently from rank and probability.  
The full light curves of about  250 candidates 
thus selected were inspected by eye. Eclipsing
variables, RR~Lyr and high-amplitude $\delta$ Sct stars were found  and not
considered furthermore. We remained with a sample where the dichotomy was again between Cepheid and
rotational variables. The stars  characterized by very large cycle-to-cycle variations were 
immediately considered as rotational variables.  The stars showing  more stable light
curves were analysed in frequency and least-squares solutions of the mean light curves were obtained.
We note that in this process the small amplitude was not considered as  
a rule of thumb since it could be reduced by an unresolved, blended star. On the other hand, 
a large amplitude (say, greater than 0.20~mag) was recognized as the signature
of a pulsational mechanism. Most of the stars were considered as rotational variables after that an
ensemble of several parameters (period, amplitude, stability of the light-curve shape, 
Fourier parameters,~...) was taken into account. However, we cannot rule out that a few
of them are actually Cepheids, whose regular curve  
is perturbed by that of a blended irregular variable. We could identify some
borderline cases 
(e.g. CoRoT 0102635101, 0102640143, 0605539616, 0223971984, 0102578739, 0110827792, and,
overall, 0223971984)
where the distinction between activity and pulsation was very difficult.
New spectroscopic observations are necessary to establish the nature of these problematic
stars in an unambiguous way.

During the screening procedure we discovered the unique Galactic triple-mode Cepheid 
CoRoT 0223989566 \citep{pbw}
and at the end we could count seven bona-fide Cepheids. In addition to
\cep, V376~Mon, and DP~Mon we identified  four new Cepheids:
CoRoT~0221640090, 0315221415, 0211626074, and 0659466327 (Table~\ref{runs}).
The CoRoT photometry of \cepdue\, and of other Cepheids was analysed when the interpretation of
the \cep\, data was almost finished. Therefore, we describe the procedure
as applied to \cep\, first and then how it has been extended to other Cepheids later.  

\begin{table*}
\caption{The Cepheids observed with CoRoT. CHR: chromatic mode, MON: white light only.
}
\begin{tabular} {rc  lrrr  lr r }
\hline
\multicolumn{1}{c}{CoRoT ID}& \multicolumn{1}{c}{Other designation} & \multicolumn{1}{c}{Run} &
\multicolumn{1}{c}{Start date} & \multicolumn{1}{c}{End date} & \multicolumn{1}{c}{$\Delta$T} & 
\multicolumn{1}{c}{Mode} & \multicolumn{1}{c}{Cadence} & \multicolumn{1}{c}{N} \\
\hline
0102618121 &          & LRa01 & Oct. 23, 2007 & Mar.  3, 2008 & 131.46 d & CHR &  32 s  & 326634 \\
0102618121 &          & LRa06 & Jan. 12, 2012 & Mar. 29, 2012 &  76.61 d & CHR &  32 s  & 196032 \\
0300003751 & V376 Mon & LRa02 & Nov. 16, 2008 & Mar.  8, 2009 & 111.64 d & CHR & 512 s  &  16400 \\
0221644967 & DP Mon   & SRa02 & Oct. 11, 2008 & Nov. 12, 2008 & 31.75 d & CHR &  32 s  &  72144 \\
0659466327 &          & LRc09 & Apr. 12, 2012 & July  5, 2012 & 83.51 d & MON & 512 s  &  11894 \\
0315221415 &          & SRa03 & Mar.  5, 2010 & Mar. 29, 2010 & 24.73 d & MON & 512 s  &   3670 \\ 
0211626074 &          & SRc01 & Apr. 13, 2007 & May   9, 2007 & 25.54 d & MON & 512 s  &   3996 \\
0221640090 & NSVS 12485452&SRa02&Oct. 11, 2008 & Nov. 12, 2008 & 31.75 d & CHR &  32 s  &  73326 \\
\hline
\label{runs}
\end{tabular}
\end{table*}

\begin{figure}
\resizebox{\hsize}{!}{\includegraphics{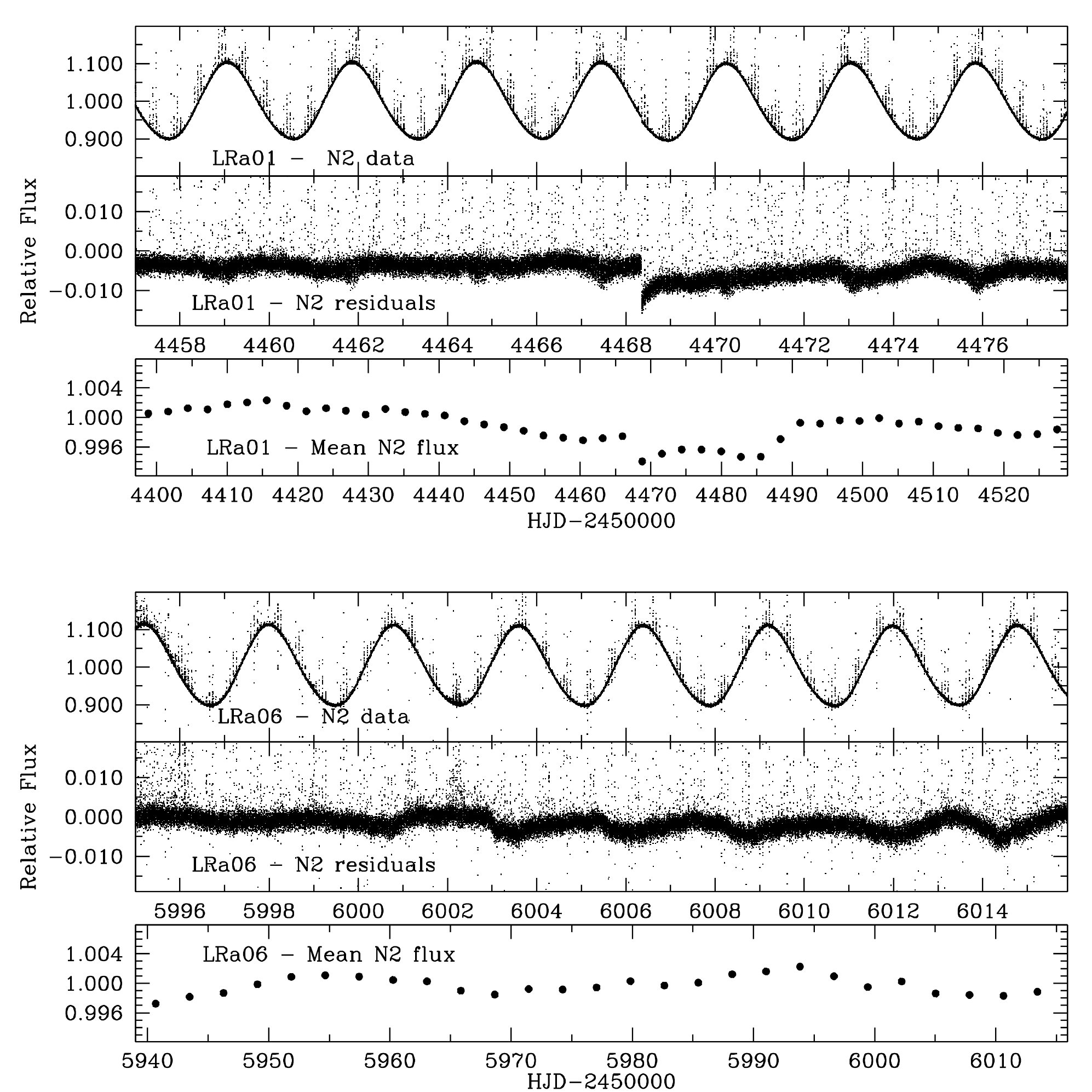}}
\caption{Frequency analysis of CoRoT N2 data of \cep. {\it Upper part}, LRa01 data:  example of light curve
(top panel), residuals after prewhitening of the main periodicity from the global dataset (middle panel), behaviour of the 
mean flux all along LRa01 (bottom panel).
{\it Lower part}: same as in the upper part, but for LRa06 data.}
\label{lruns}
\end{figure}

\section{The analysis of \cep\, data}
\subsection{Spectroscopic follow-up}
To solve the ambiguous spectral classification of \cep\, and to validate our classification criteria, 
we obtained three high-resolution spectra  on 2012 January 10 and 11 with the HARPS spectrograph 
installed at the 3.6-m telescope of the ESO-La Silla Observatory. 
The spectra were taken in the high-efficiency EGGS configuration (R=80,000) and they have signal-to-noise ratios (S/N) between 50 and 55.
We reduced and normalized the spectra with a semi-automated pipeline developed at the Brera Observatory 
\citep{laurea}.
We then computed the mean line profiles for each spectrum using the LSD software \citep{lsd}
and we used them to estimate the radial (from 47.9 to 53.6~\kms) and projected rotational 
(from 9.64 to 11.85~\kms) velocities by fitting the profiles with gaussians. 
The reduced and normalized spectra were shifted by their radial velocities and averaged in order to 
lessen the effect of the pulsation.
The spectral synthesis software SME version 4.2.3 \citep{sme}
was used to estimate the physical parameters. The spectral regions 5160-5190~\AA\, 
(including the gravity sensitive Mg-triplet) and 4830-4900 \AA\, (the H$\beta$ region, 
sensitive to the temperature) were selected. We left temperature (\teff), gravity ($\log g$, cgs units), 
metallicity ([Fe/H]), micro- and macroturbulence as free parameters, while $v\sin i=11~$\kms\, was 
imposed.  Following \citet{valenti},
we also left the abundances of Na, Si, Ti, Fe, and Ni as free parameters.
We performed an iterative process on the MARCS stellar models \citep{marcs}: 
we started running the SME software on the Mg region, then we 
switched to the H$\beta$ region keeping the $\log g$ previously found as a fixed parameter. 
After that we went back to the Mg region keeping fixed the \teff\, value found in the H$\beta$  region, 
and so on until the results converged. 
The resulting parameters are \teff= 6087$\pm$44\,K,  $\log g$=1.53$\pm$0.15~dex, [Fe/H]=$-0.34\pm0.04$~dex. 
They confirm that \cep\, is a Pop.~I supergiant and not a main-sequence star: its classification as a Cepheid
variable was definitely established.
\subsection{CoRoT data}
 The original timeseries of \cep\,  consist of 348~702 and 206~528 data points, 
respectively. In both cases the exposure time was
set to 32~s (corresponding to the CoRoT short cadence) and the flux was recorded
in three different spectral regions, i.e., in the CoRoT chromatic mode. 
The contamination factors  (in a scale ranging from 0 to 1) are 0.026 and 0.036 in the two CoRoT runs,
suggesting no relevant contribution from back- or fore-ground stars.

The analysis of the original N2 data clearly
detected a predominant peak at \fu=0.3574~\cd\, followed by several harmonics. 
Figure~\ref{lruns} shows LRa01 (top panel of the upper part) and LRa06
(top panel of the lower part) CoRoT N2 data on a 15-d time baseline. All data points 
are plotted, including those acquired during the crossing of the South Atlantic Anomaly
(SAA): they  appear as bright outliers on a vertical line. We avoided any cleaning
procedure since 
the detection of instrumental trends and jumps and their subsequent correction
is a crucial step to proceed in the analysis of the CoRoT data. 
Therefore, we computed a preliminary solution based on the set of frequencies \fu, 2\fu, 3\fu, 
4\fu, and 5\fu. The  plots of the residuals (middle panels of the two parts of  Fig.~\ref{lruns})
show sudden jumps around JD 2454468 in LRa01 and JD 2456003  in LRa06, the SAA outliers,
and continuous oscillations. Outliers were iteratively removed at this stage.
We still remained with the necessity to correct the
sudden jumps  and long-term drifts and to verify the origin of the residual oscillations.
To do this, we applied the least-squares fits to each pulsation cycle, still using the
set of frequencies from \fu\, to 5\fu. 
The 32-s cadence of the CoRoT data warranted more than 7000~measurements for each cycle,
making the solution very robust.
This procedure returned
the mean fluxes of the 47 cycles observed in the LRa01 and the 27 ones in LRa06 (Fig.~\ref{lruns}, 
bottom panels of the two parts).
These mean values show the variations due to the jumps and the long-term trend.
We notice that the long-term trends behave in different ways in the two runs.
 The effect of the jump around JD~2454468 was noticeable in the data until 
JD~2454485, thus affecting the mean magnitudes of the  subsequent six cycles.
To remove the instrumental variations, we interpolated the values
of the mean fluxes (bottom panels of Fig.\ref{lruns}) at the times of each individual measurement and 
then we subtracted it.  At this point  
two cleaned timeseries (326,634 and 196,032 points, respectively) were ready 
to be analysed in frequency.
By applying this procedure we probably modified any possible additional 
signal around 2.8~d and this fact has to be carefully considered when discussing the
frequency content of the light curves of the two runs.

We transformed the fluxes  into magnitudes by calculating the average flux of each run.
We also measured the level of the noise of the CoRoT timeseries after subtracting the set of
the \fu\, frequency and harmonics. In the case of the LRa01 data we obtained 0.065~mmag in the 0.01-1.0~\cd\, range,
rapidly decreasing to 0.018~mmag in the 1.0-2.0~\cd\, region, and more slowly to 0.012~mmag
in the 2.1-3.0~\cd\, region and to 0.007~mmag around 5.0~\cd. In the case of the LRa06
data the four  values are 0.104, 0.030, 0.020, and 0.010~mmag, respectively.
The number of measurements (326,634 versus 196,032) accounts for most of the difference between the noise levels
of the LRa01 and LRa06 spectra.

\section{A critical analysis of the frequency content of \cep}
\begin{figure}
\resizebox{\hsize}{!}{\includegraphics{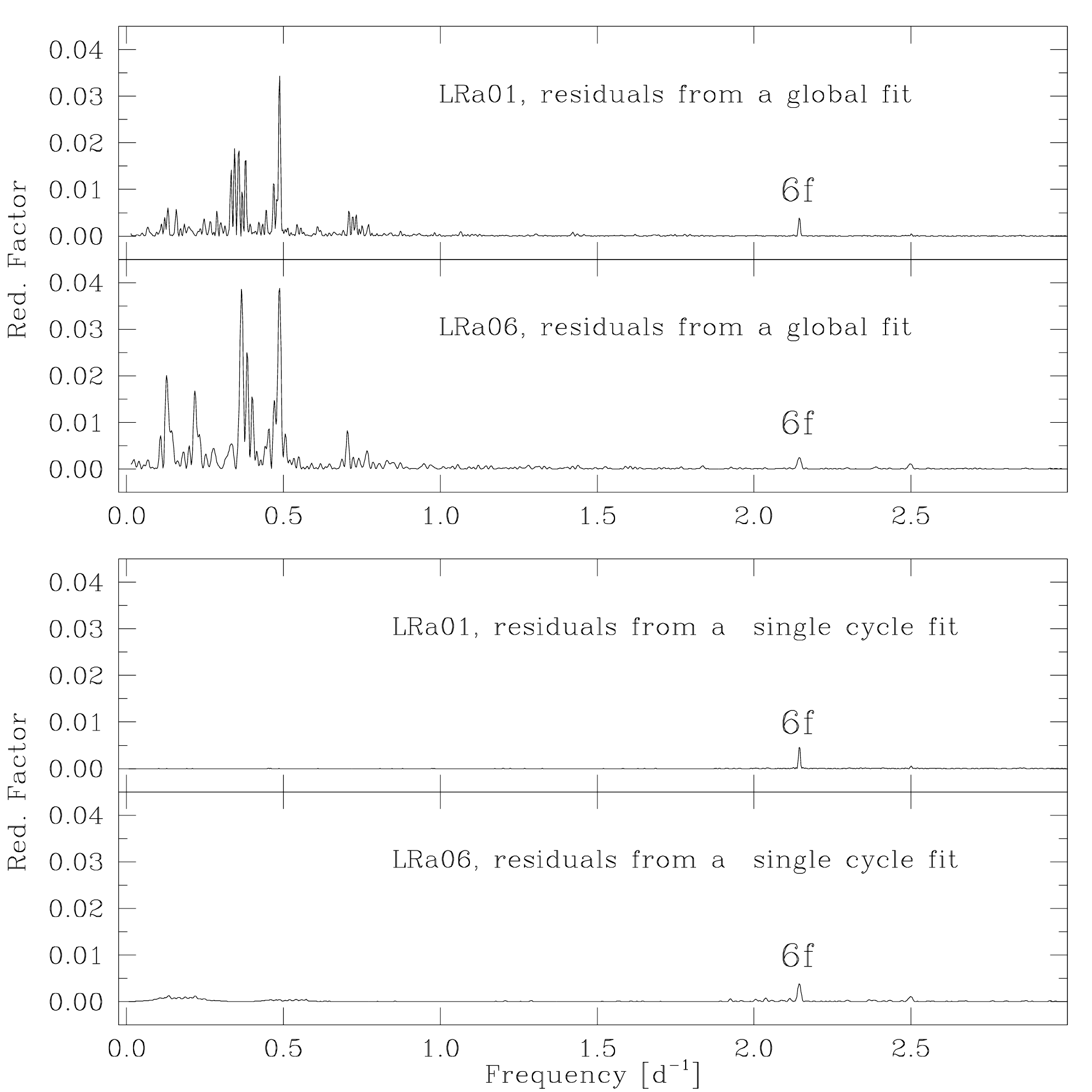}}
\caption{Analysis of CoRoT N2 data of \cep. {\it Upper part},  
power spectra of the residuals after applying a global fit to the entire set
of  the LRa01 data (top panel) and LRa06 data (bottom panel).
{\it Lower part}: power spectra of the residuals after applying a fit
to each pulsational cycle of the  LRa01 data (top panel) and LRa06 data (bottom panel).}
\label{spectrares}
\end{figure}
\begin{figure}
\resizebox{\hsize}{!}{\includegraphics{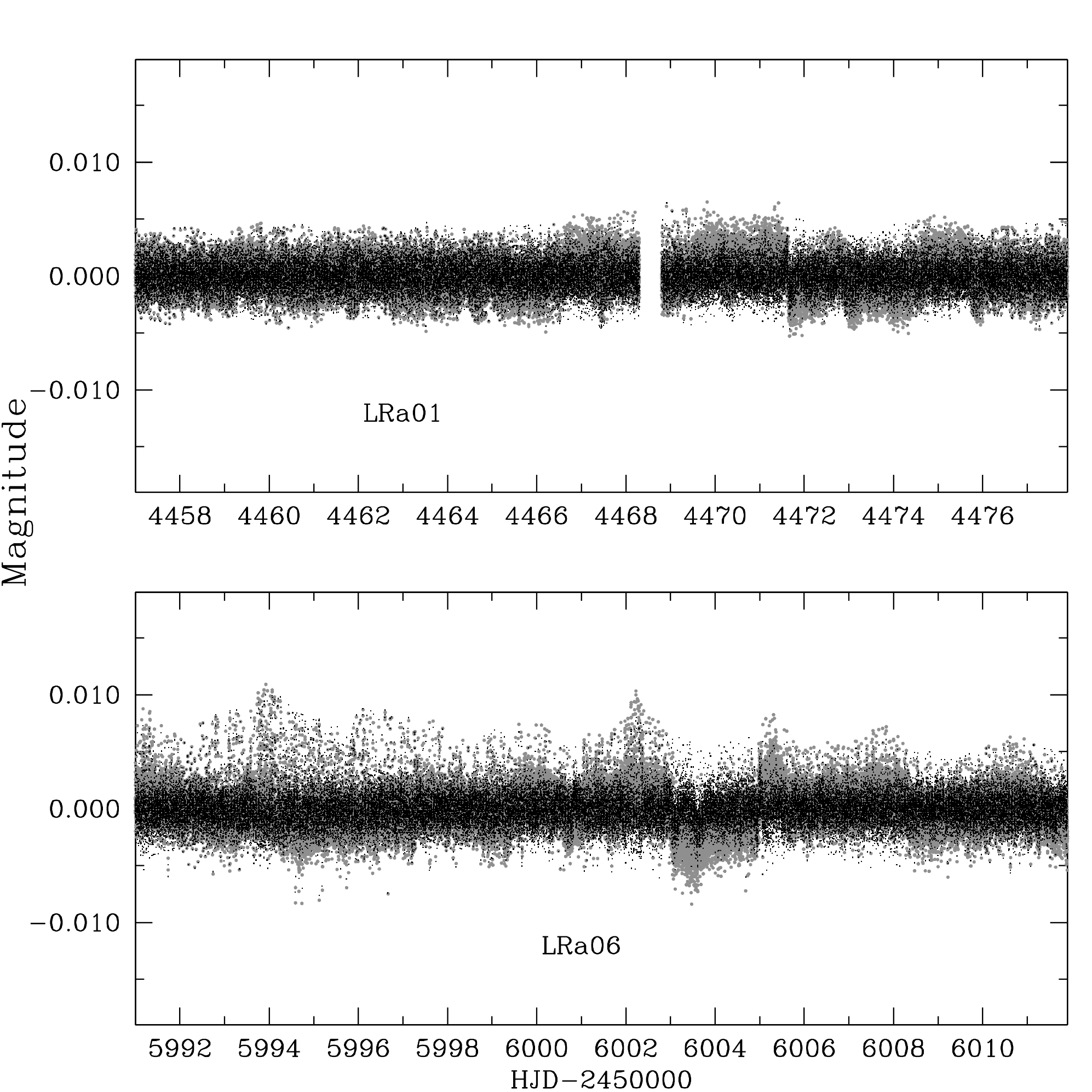}}
\caption{
{\it Upper part}, LRa01 data:  
light curves of the residuals after subtracting \fu, 2\fu, 3\fu, 4\fu, and 5\fu\, 
calculated from the entire set of data (grey points) and from each pulsational cycle (black points).
A part of the data is shown.
{\it Lower part}, the same as in upper part, but for LRa06 data.}
\label{curveres}
\end{figure}
The frequency analysis was performed by means of the iterative sine-wave, least-squares
fitting method \citep{vani} and then refined with the MTRAP algorithm \citep{mtrap}, allowing
us to include all the significant harmonics in the search for the best fit around \fu. 
The best way to illustrate our final results is to follow the chronological order of their
progressive consolidation.  

The analysis of the LRa01 data in white light was concluded in 2011 and the results were
apparently very promising. We found the harmonics up to 6\fu\, (amplitude 0.13~mmag),
flanked by some modulation terms
at $\pm$0.007~\cd\, (i.e., 142~d). Since the separation is very close to the length of
the observations, the origin of these terms has to be ascribed to a residual, still 
imperfect correction of the instrumental drifts.
Moreover, a few peaks not very close to \fu\, and 2\fu\,  were detected: 
0.488~\cd\, (amplitude 0.40 mmag), 0.392~\cd\, (0.14~mmag), and
0.288~\cd\, (0.15~mmag). Only the peak at 0.488~\cd\, had 
a S/N above the threshold considered as the significativity one in the Fourier analysis
\citep[S/N=4.0;][]{snr4}, i.e., S/N=0.396/0.065=6.1.
The power spectra illustrating the bunch of low amplitude peaks still centred around \fu,
the 0.488-\cd\, peak,  and the 6\fu\, harmonic are shown in the top panel of  Fig.~\ref{spectrares}. 
However, the identification of the 0.488~\cd\, peak as an additional  mode  was
not really convincing because no combination term was observed.

Why this non-detection is  so important?  The combination or coupling  terms  
are generally  ascribed  to the non-linear responses of  the stellar atmospheres to the strong 
radial pulsations. The frequency analysis of the light curves of double-mode Cepheids 
show several of them, also involving harmonics of the $F$ and $1O$
radial modes \citep{pardo}. The Fourier decomposition of the light curves also shows
regularities in the distribution of the phase parameters that are not yet fully 
understood \citep{ccterm}. The combination terms have been detected in the light curves
of other multimode pulsators, e.g.  HADS stars \citep{gsc}, RR~Lyr stars \citep{gruber},
and white dwarfs \citep{wu}.
The CoRoT photometry also allowed us to detect
low-amplitude, nonradial modes in an HADS star pulsating in a single radial mode 
and also in this case a lot of  combination terms
were detected \citep{hads}. Therefore, we should expect combination terms when  additional 
modes, both radial and nonradial, are excited in a high-amplitude radial pulsator.

The uncertainties on the real significativity of the peaks detected in the LRa01 data
slowed down our theoretical analysis and made us unsure of how to proceed further. 
The re-observation of the same field in the LRa06 pointing supplied us
with new data to verify the frequency content of \cep.
We applied the same procedure used for the LRa01 data to remove the content of the main
variation.  The frequency analysis of the residuals of the  LRa06 data confirmed 
the presence of 
the 0.488~\cd\, peak (Fig.~\ref{spectrares}, bottom panel of the upper part).
Therefore, the confidence on the stellar origin of the 0.488~\cd\, grew.
However, the ratio 0.357/0.488=0.732 would be quite unusual for a double-mode
Cepheid with a normal metallic content, both from observational data \citep{pardo,kate} and 
theoretical models \citep{bszabo,buchler}. Therefore, such a ratio
pointed out a possible nonradial mode. The identification of the peak as the rotational
frequency is much more unlikely, since the corresponding rotational period (2.06~d) appears
too short for a supergiant star. 

The excitation of nonradial modes in a classical Cepheid
would be  a very important ingredient in the  modelling of the atmospheres of
these supergiant stars, thus we submitted this result to further checks.
At this stage, we detected the 0.488~\cd\, peak in the residuals of each Long Run.
These sets of residuals were obtained by applying a global fit to the LRa01 data 
(grey points in the upper panel of Fig.~\ref{curveres}) and to the LRa06 ones
(grey points in the lower panel of Fig.~\ref{curveres}).
Indeed, variations in the shape of the residuals are clearly visible,
especially after the jumps at JD~2454468 and JD ~2456003.
They remind the oscillations noticed before removing the long-term trends 
and the jumps (Fig.~\ref{lruns}, middle panels).
This suggested that the computation of the mean fluxes was very effective
in cleaning the power spectra,
but the interplay betweeen the large amplitude of \cep\, and the
jump correction could somehow leave spurious variability  close to the significance level.
For this reason we gave a closer look to the light curves of the residuals from which the 0.488~\cd\, peak was
detected.  These light curves show the same type of oscillations,
suggesting that the global solution was not able to match 
the instrumental alterations of the stellar signal.
Thus we decided to calculate residuals following the same 
approach used to determine the mean fluxes.
This implied to compute as many sets of parameters as the elapsed cycles were, i.e.,
47 for LRa01 and 27 for LRa06.

The analysis of these new sets of residuals were enlightening.
The oscillations visible in Fig.~\ref{lruns} (middle panels) had completely 
disappeared and the light curves became really flat, with
just some clues of high-frequency noise (Fig.~\ref{curveres}, 
black points).  Indeed, the power spectra of the 
residuals from fitting one cycle at a time show the peak at the 6\fu\, harmonic only.
This was observed both in the LRa01 data  
and in the  LRa06 data (lower part of Fig.~\ref{spectrares}).
Both spectra are characterized by an almost white noise.
It could be argued that the signal at 0.488~\cd\, was removed  when fitting the single cycles  
(also taking into account the almost exact 3/4 ratio between the frequencies). However, the 
fact that the 6\fu\, peak has remained clearly visible gave evidence that the
fit with \fu, 2\fu, 3\fu, 4\fu, and 5\fu\, of each cycle did not modified any signal component
apart from these frequencies.
We tried to come back to a new correction of the
mean fluxes, but again the analysis of the global residuals thus obtained was
unsatisfactory.
We note that the technique of analysing single cycles  were
successfully applied to remove  similar instrumental effects in the light curves of  
the high-amplitude $\delta$ Sct star CoRoT~101155310 \citep{hads} 
and of the triple-mode Cepheid CoRoT 0223989566 \citep{pbw}. 

\begin{figure}
\resizebox{\hsize}{!}{\includegraphics{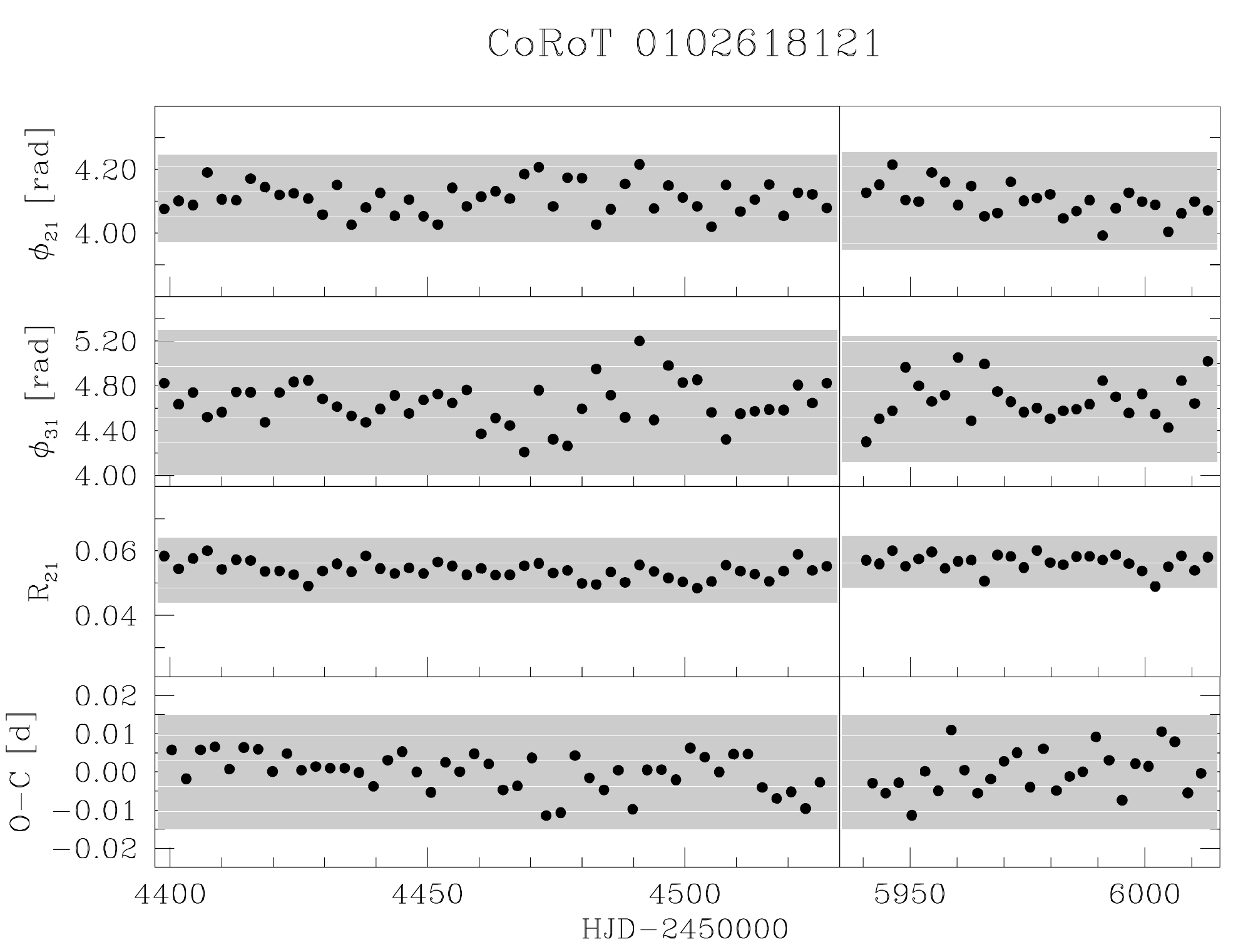}}
\caption{The Fourier parameters and O-C values as measurements of the cycle-to-cycle variations in the \cep\,
light curve.  The grey boxes are the $\pm3\sigma$ bands around the average values. {\it Left part:} LRa01 data;
{\it Right part:} LRa06 data.
} 
\label{fourier}
\end{figure}

\begin{table*}
\caption{Fourier parameters of
the curves  of \cep\, in the different colours. }
\begin{tabular} {l  llll c llll}
\hline
\multicolumn{1}{c}{Term}&\multicolumn{4}{c}{LRa01} &&\multicolumn{4}{c}{LRa06}\\
\cline{2-5} \cline{7-10}
\multicolumn{1}{c}{}  & \multicolumn{1}{c}{white} &\multicolumn{1}{c}{\rc} &
\multicolumn{1}{c}{\gc} & \multicolumn{1}{c}{\bc} && \multicolumn{1}{c}{white}&
\multicolumn{1}{c}{\rc} & \multicolumn{1}{c}{\gc} & \multicolumn{1}{c}{\bc} \\
\fdu\,[rad] & 4.1101(7) &  4.220(2)   &  4.070(2) &    3.934(2)  &&   4.1081(9) & 4.191(2)  & 4.012(6)    & 3.925(6)\\
\ftu\,[rad] & 4.645(6)  &  4.637(13)  &  4.499(21) &   4.683(20) &&   4.689(7)  & 4.674(15) & 4.814(84)   & 4.920(64)\\
\rdu & 0.05395(4)&  0.05371(9) &  0.05260(12) & 0.05166(11) && 0.05646(5)& 0.05717(14)& 0.05265(34)& 0.04961(30)\\
$\phi_{41}$\,[rad] &  3.624(7)& 3.857(18) & 3.431(21)& 3.240(17) &&  3.573(7) &  3.712(26) &3.290(34)  & 3.193(32)\\
\noalign{\smallskip}
Full ampl. [mag]  & 0.222 &  0.191 &  0.250 & 0.306& &    0.233 &0.208 & 0.288 & 0.364\\
\noalign{\smallskip}
Res. rms [mag]& 0.0017 & 0.0036 & 0.0061 & 0.0063 && 0.0018 & 0.0045 & 0.0150 & 0.0165 \\
\noalign{\smallskip}
\hline
\label{colori}
\end{tabular}
\end{table*}

\section{The properties of the \cep\, light curves}
Once that we could rule out the excitation of other modes, 
we fitted the CoRoT magnitudes by means of a cosine series
(see appendix~\ref{appen}).
Table~\ref{sol} lists the coefficients  of the solutions of the light curves of
both runs. The shape of the light curve could be defined by the Fourier parameters
(see appendix~\ref{appen}) and 
Table~\ref{colori} lists those relevant to our discussion.
\subsection{The stability of the light curves}
The absence of any relevant peak once that the contribution of the main
oscillation has been removed is not by itself an evidence that the light
curve is stable. Cycle-to-cycle variations as those observed in the {\it Kepler}
light curve of V1154 Cyg or in the radial velocity curves could not result in 
a well defined peak in the power spectrum. Moreover, at least in principle, 
the residual peaks left around \fu\, when applying
a global fit could be ascribed to cycle-to-cycle variations, maybe erratic 
ones. 

We analysed the \rdu, \fdu, and \ftu\, parameters of each cycle 
to quantify how much the light curve deviates from the average behaviour.
The CoRoT 32-s cadence allowed us to get about 7500 measurements per cycle,
adequate to survey the harmonics 2\fu\, and 3\fu\, in a very accurate way, though their amplitudes are
very small (6.0 and 0.7~mmag, respectively). 
The Fourier parameters are shown in Fig.~\ref{fourier}. The error bars 
are three times the standard deviations ($\pm3\sigma$) around
the average of the parameters of the individual cycles.
The immediate result is that the differences from the average values are small.
We could notice that all the individual values are between
3$\sigma$, though some apparently regular variations are sometimes visible (e.g.
the \rdu\, behaviour in the LRa01). 
Therefore, we performed a frequency analysis of the Fourier parameters.
We could not detect any periodicity since 
the power spectra of the \rdu, \fdu, and \ftu\, parameters
do not show any common feature either in a single Long Run or when comparing LRa01 and LRa06 results. 

\begin{figure}
\resizebox{\hsize}{!}{\includegraphics{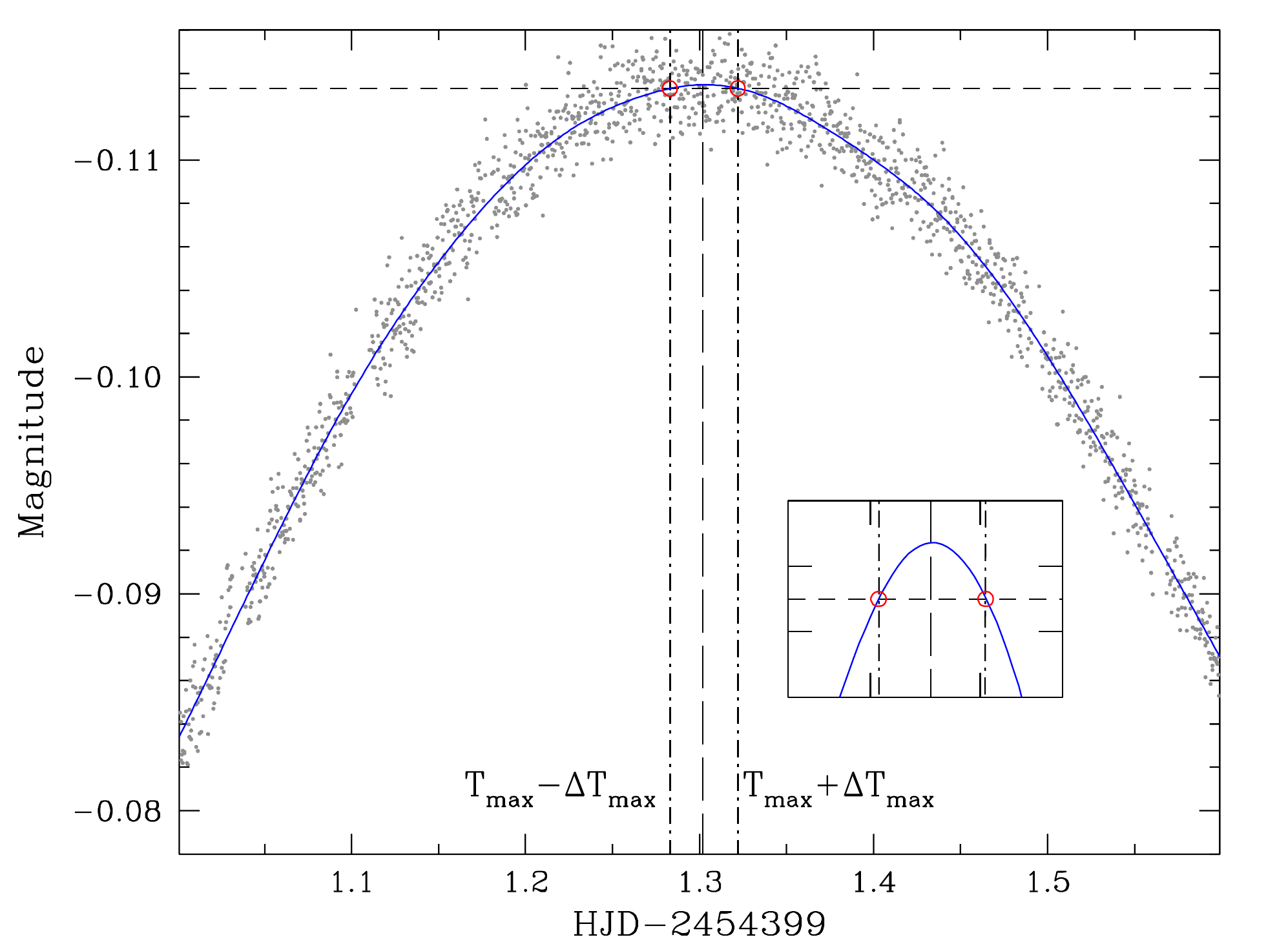}}
\caption{Determination of the error bar $\Delta$\tmax\, of the first \tmax\, of \cep. 
The original data are in grey, the 
cubic spline fit is the blue continuous line, the $y=m_{max} + \sigma/\sqrt{n-1}$ the horizontal 
dashed line; the two red circles indicate the intersection between the
spline curve and the line (better shown in the inset). The differences of their abscissae with 
respect to the \tmax\, one fix the error bars $\Delta$\tmax.}
\label{spline}
\end{figure}

However, the best parameter to quantify the stability of the light curve
is probably the O-C parameter.
We determined the observed \tmax 's 
by means of  cubic spline functions with a non-zero smoothing parameter that
depends on the number of measurements to fit and on their scatter
\citep{reinsch}. We thus obtained the linear ephemeris from the least-squares
method
$
\begin{array}{lrrr}
{\rm T_{\it max} =  HJD}& 2454400.2977 & +~ 2.798061&\cdot~{\rm E}\\
                &    \pm0.0008 &~\pm0.000002&  
\end{array}
$\\
(E$=0,1,2,..., n$) 
and hence we could calculate the O-C value for each \tmax.
The residual rms of the least-squares linear fit is 0.0052~d and the 3$\sigma$ band is shown
in Fig.~\ref{fourier}. We also estimated the uncertainty on the individual \tmax\,
determination as the difference between the two times corresponding to
the intersection of the spline function with the line $y=m_{max} + \sigma/\sqrt{n-1}$ 
(Fig.~\ref{spline}),
where $m_{max}$ is the magnitude at maximum and $\sigma$ is the standard deviation
of the fit. Such uncertainties are ranging from 0.010 to 0.016~d and hence the 
3$\sigma$ interval (i.e., 22.5~min) matches the individual error bars, too. 
We analysed in frequency the O-C values too, again obtaining no clear indication of periodicities.  

As a final conclusion, looking at the different plots shown in Fig.~\ref{fourier}
and at the results of the various frequency analyses, we cannot identify any additional
periodicity in the variability of \cep. The Fourier parameters and the O-C values
sometimes mimic an apparent regular behaviour. This could be a hint of cycle-to-cycle
variations, much less regular than those observed 
in the {\it Kepler} light curve of V1154~Cyg \citep{poster}. 

\subsection{The comparison between LRa01 and LRa06 light curves}
We merged the LRa01 and LRa06 measurements and we calculated a refined value
of the period on the whole time baseline. The MTRAP code supplied the final
vale $P$=2.7980596$\pm$0.0000001~d, in excellent agreement with the value
obtained from the \tmax 's values only.
The most relevant difference in the datasets of the two Long Runs is 
a slight increase from 0.222 to 0.233~mag in the full-amplitude of the light curves.
Since we are dealing with absolute photometry, i.e., without any comparison star, we have to
be sure that the measurements were performed in the same instrumental system to
infer that the amplitude difference has a stellar origin. We immediately noticed that the star was
observed in the same CCD in the two runs, but with a different mask and at different pixel
coordinates. As a results, the average white 
flux measured was of 1~030~069~electrons in LRa01 and 944,911~electrons in LRa06.
Moreover, the flux subdvision in \rc, \gc, \bc\, colours was also different: 65, 14,
21 per cent  in LRa01, 78, 10, 12 per cent in LRa06. We conclude that the small difference in amplitude
could very probably be ascribed to the different settings of the CoRoT pointings in the two runs.
The same effect occurs in {\it Kepler} photometry: average fluxes and amplitudes
change for the same star observed 
in the same CCD but with different masks.  These effects 
are particularly disturbing when dealing with 
large amplitude variables \citep{benko}.

The procedure applied to  white light measurements was repeated for
each colour of both runs.  Other similar, subtle  instrumental effects became  evident
during this process \citep{kasc}. Then, the Fourier decomposition was applied to the cleaned timeseries
in each colour (Table~\ref{colori}). The parameters clearly show the expected dependence from the bandpass
\citep[see ][for a comparison between $V$ and $I$ parameters]{morgan}.  The different flux  distributions
between LRa01 and LRa06 also account for the small differences between parameters in the same colour.

\section{Data analysis of the other Cepheids observed with CoRoT}
\begin{figure*}
\resizebox{\hsize}{!}{\includegraphics{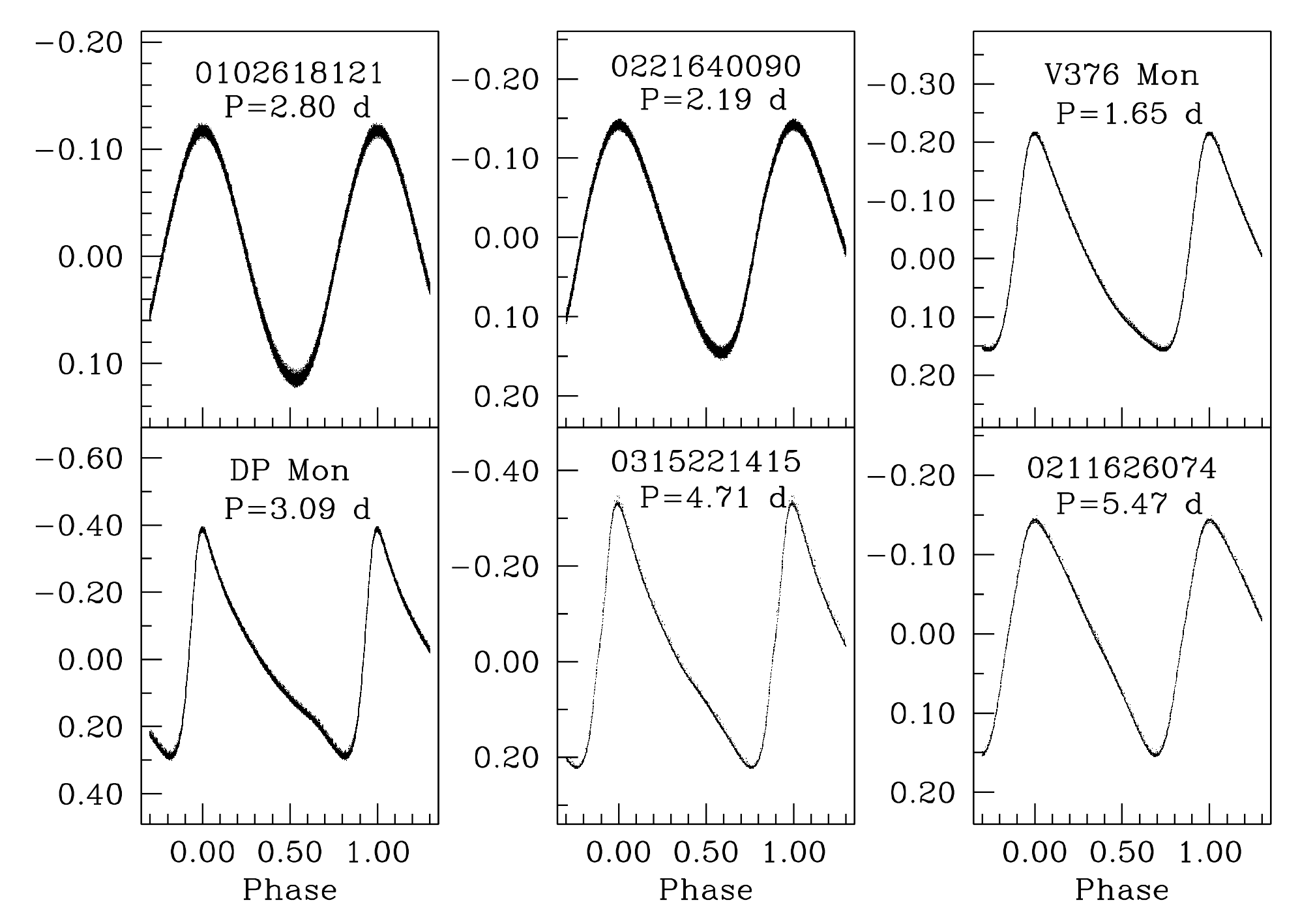}}
\caption{Family portrait of the CoRoT light curves of Pop.~I\, Cepheids.
}
\label{curve}
\end{figure*}

\begin{figure}
\resizebox{\hsize}{!}{\includegraphics{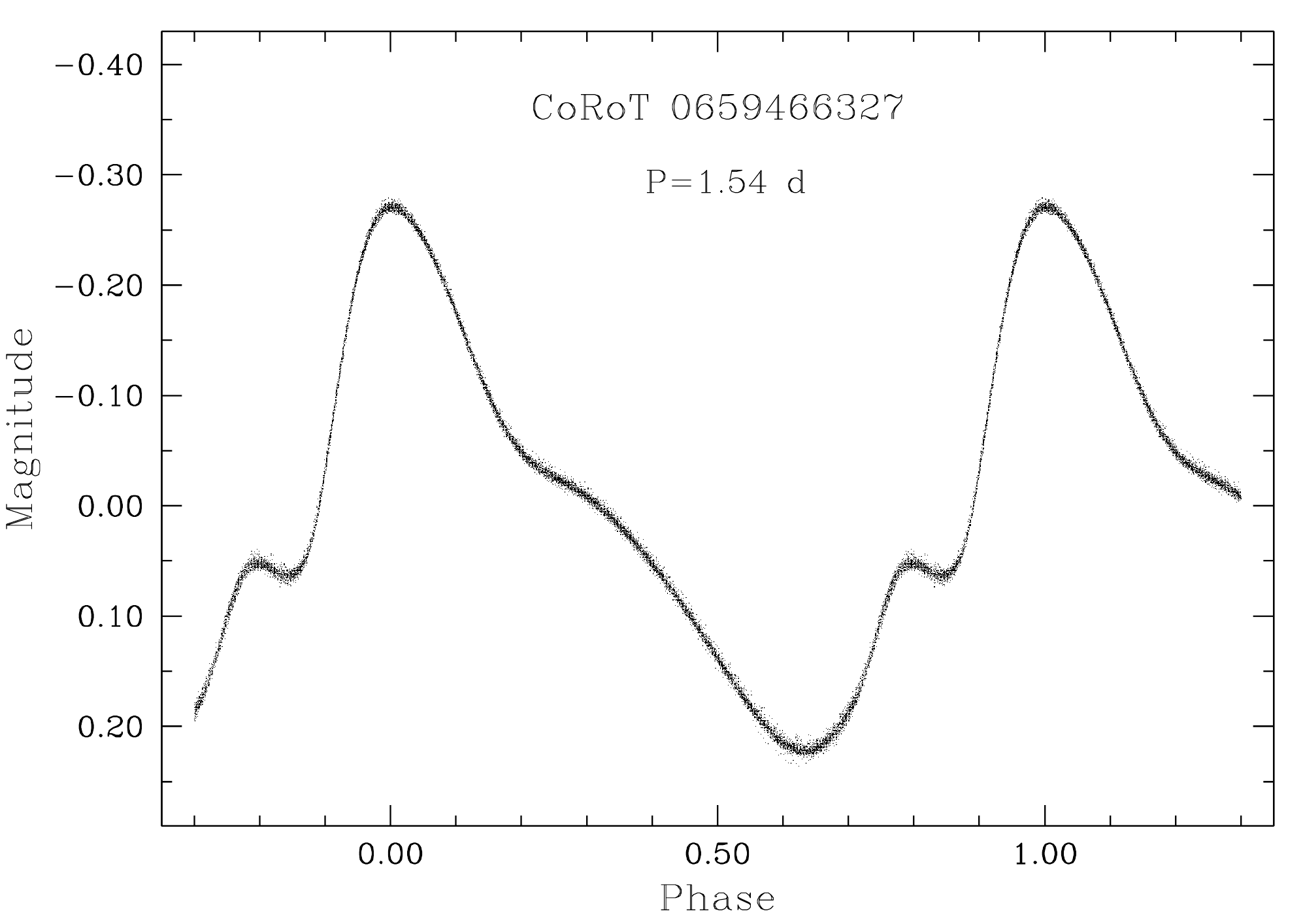}}
\caption{The light curve of the Pop.~II Cepheid CoRoT 0659466327.
}
\label{curvaii}
\end{figure}

\begin{figure}
\resizebox{\hsize}{!}{\includegraphics{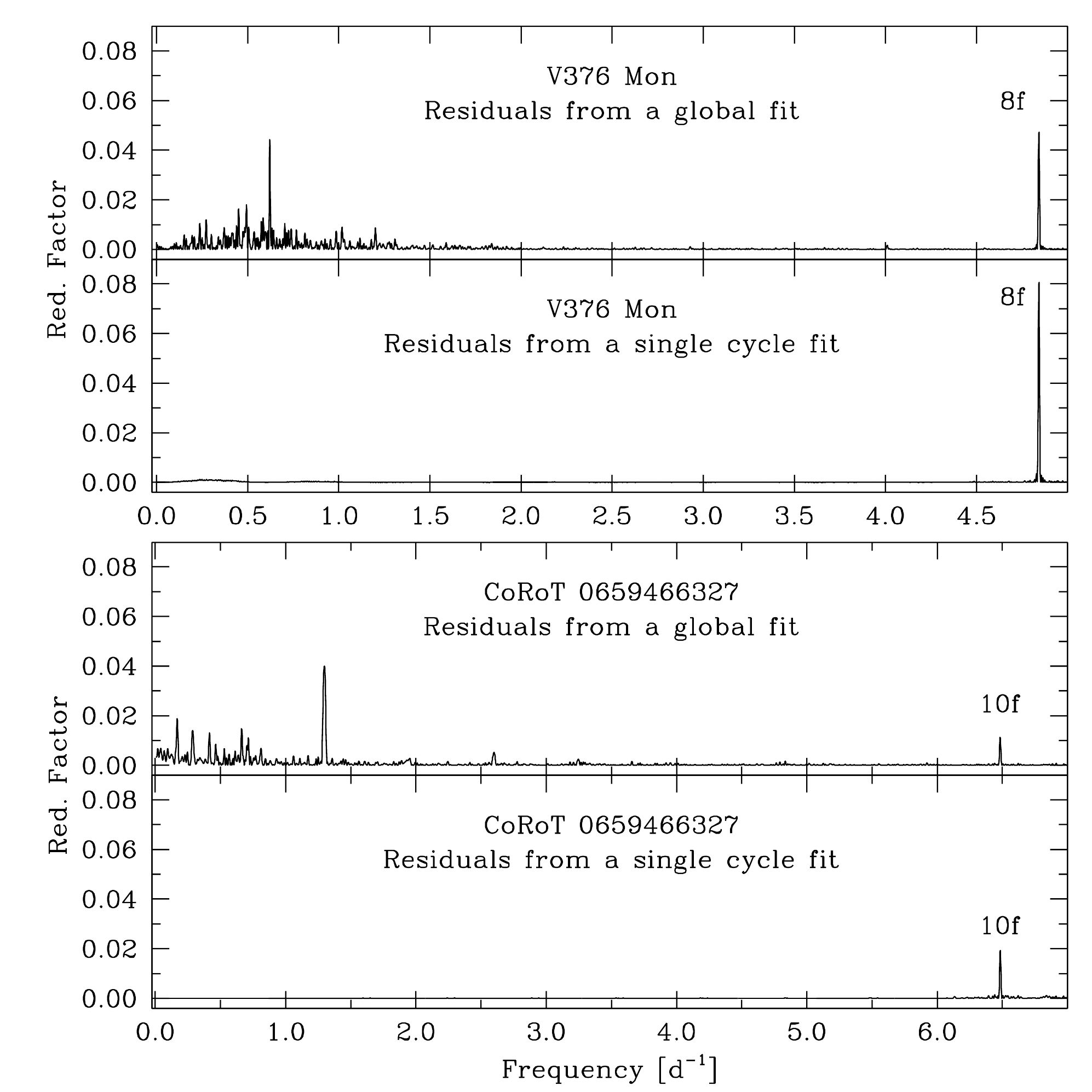}}
\caption{{\it Upper part}: analysis of CoRoT N2 data of \cepdue:  power spectra
of the residuals after subtracting a global fit up to 7\fu\, from the entire set of data (top panel),
and a single fit from each pulsational cycle (bottom panel).
{\it Lower part}: the same analysis for CoRoT 0659466327, with a fit up to 9\fu.
}
\label{cef751}
\end{figure}

The procedure described in the previous sections for \cep\, was applied
to the other Cepheids, whose folded light curves are shown in Fig.~\ref{curve}
(six Pop.~I stars) and Fig.~\ref{curvaii} (one Pop.~II star).
The Fourier parameters are listed in Table~\ref{others}.

\subsection {Stars observed in the Long Runs}
\begin{figure}
\resizebox{\hsize}{!}{\includegraphics{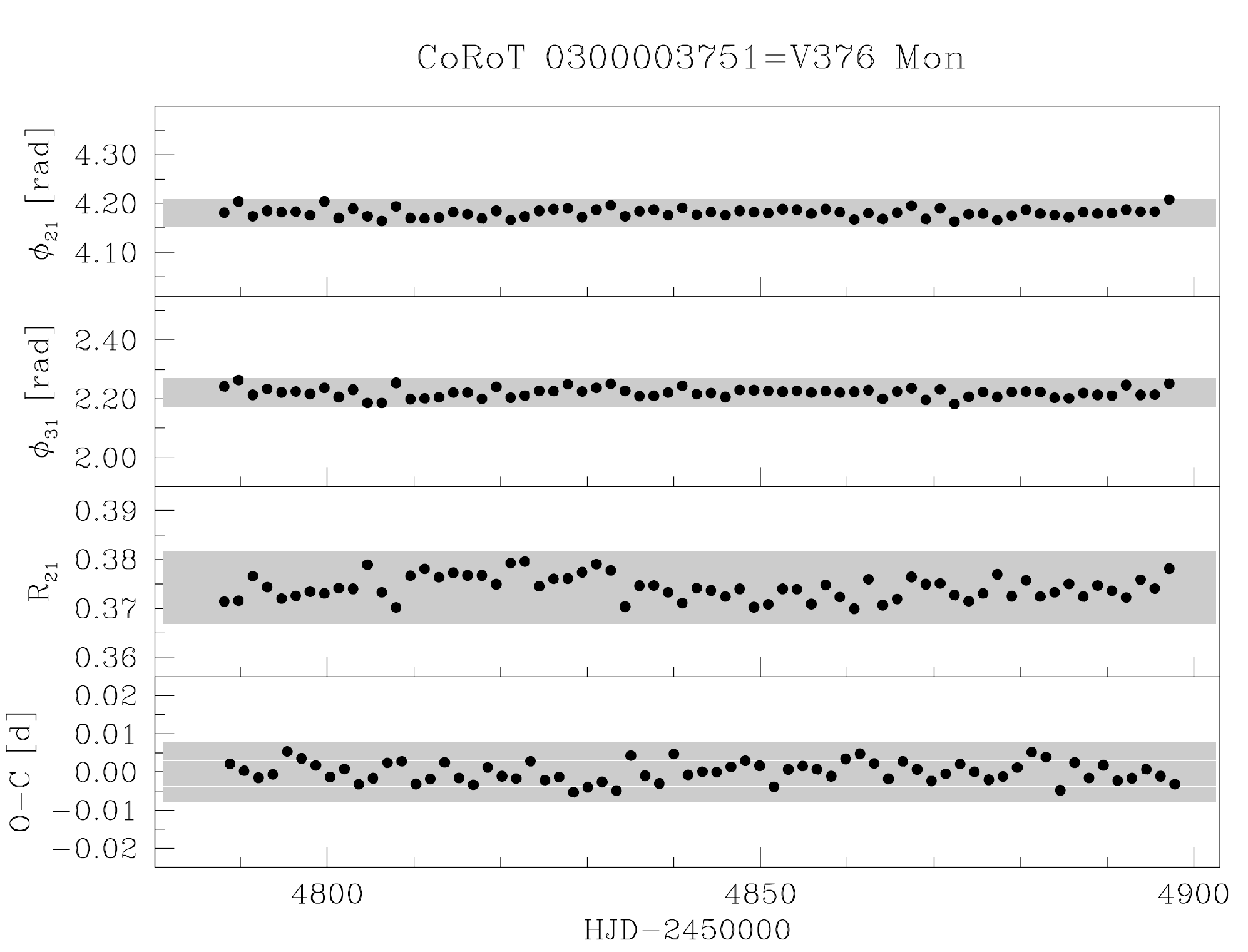}}
\caption{
The Fourier parameters and O-C values as measurements of the cycle-to-cycle variations in the \cepdue\,
light curve.
The grey boxes are the $\pm3\sigma$ bands around the average values.
}
\label{decocef751}
\end{figure}

\begin{figure}
\resizebox{\hsize}{!}{\includegraphics{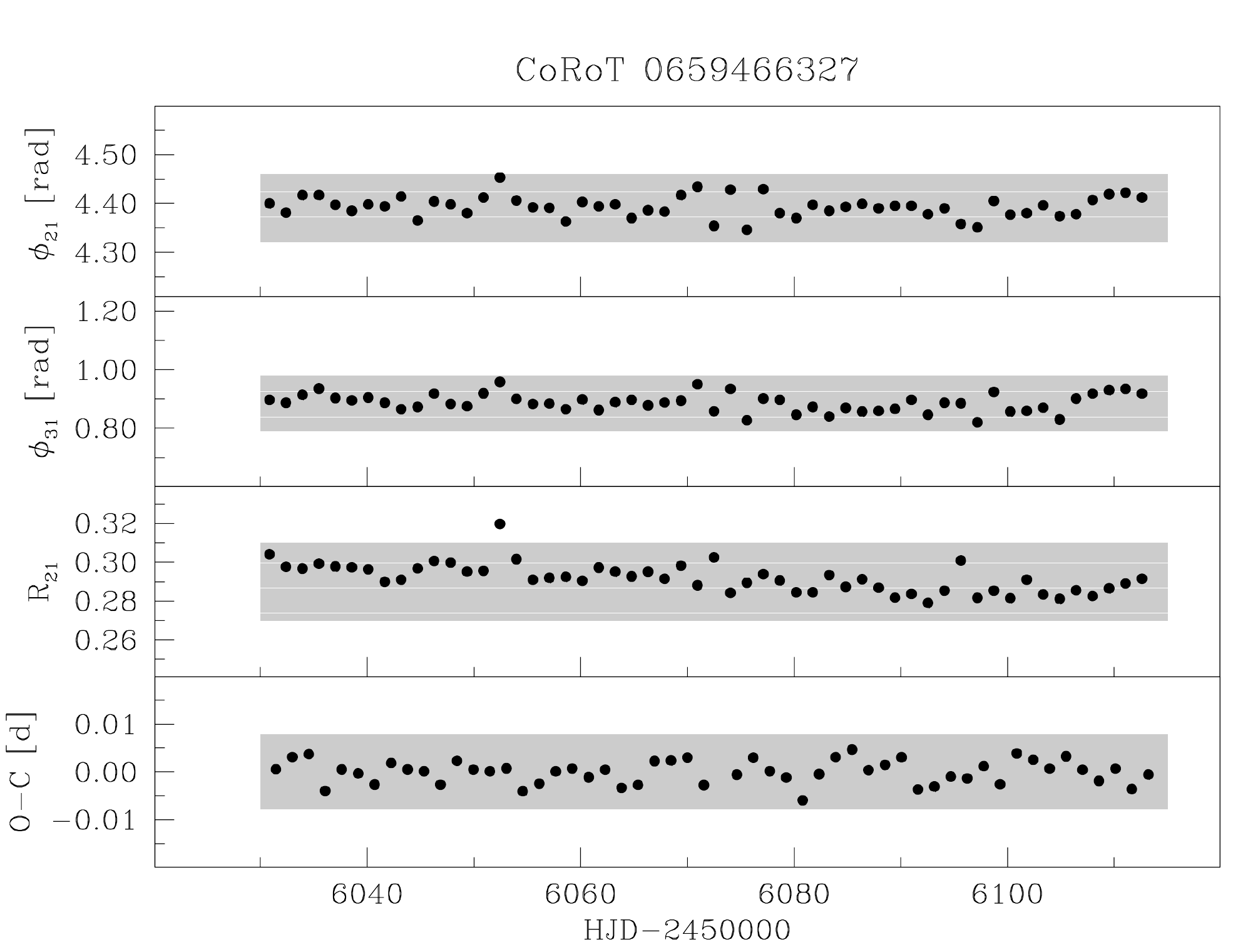}}
\caption{
The Fourier parameters and O-C values as measurements of the cycle-to-cycle variations in the 
CoRoT 0659466327 light curve.
The grey boxes are the $\pm3\sigma$ bands around the average values.
}
\label{decocef327}
\end{figure}

The frequency analysis of the \cepdue\, data
supplied \fu=0.6060~\cd, refined to  
$P=1.651969\pm0.000001$~d with MTRAP. Table~\ref{sol} lists the least-squares solution
of the CoRoT white light curve. The frequency analysis did not detect any
significant additional mode, but again the residual power spectra strongly 
depend on the procedure used to prewhiten the data. We obtained a bunch of peaks around
\fu\, when applying  a global fit up to 7\fu\, to the entire dataset. These peaks  disappeared when
the fit was calculated cycle-by-cycle (Fig.~\ref{cef751}, upper part). The harmonic 8\fu\, 
was detected in both cases, confirming that the single cycle fit did not
alter the frequency content. The Fourier parameters and the O-C values show a small scatter 
around the means (Fig.~\ref{decocef751}), suggesting no periodic variation. 

The frequency analysis of CoRoT 0659466327 data  returned \fu=0.648~\cd\, and the period was refined to
$P=1.542353\pm0.000002$~d. 
Table~\ref{solii} lists the least-squares solution of the  white light curve.
The bump on the ascending branch (Fig.~\ref{curvaii}) 
is typical of Pop.~II Cepheids.  Differently than in
the previous cases, the peak in the power spectrum of the residuals from the global fit
up to 9\fu\, occurred around 2\fu, but 
the bunch of peaks at \fu\, were detected, too. The power spectrum of the residuals from
the cycle-by-cycle fits is completely flat and it just shows the 10\fu\, harmonic,
not prewhitened from the data (Fig.~\ref{cef751}, lower part). Small-size scale variations are 
observed in the Fourier  parameters and O-C values, but again without any trace of periodicity 
(Fig.~\ref{decocef327}).

The power spectra of the residuals of \cep, \cepdue\, and CoRoT 0659466327 
calculated by means of a global least-squares
solution show the limits of this approach when applied to continuous and high-precision timeseries where
systematic errors (jumps and/or drifts) can occur. When possible, e.g. in presence of a high-amplitude
predominant mode, the frequency analysis of cycle-to-cycle residuals are preferable, taking care to 
keep a known frequency, e.g. an harmonic, as internal check of the signal preservation.

\subsection{Stars observed in the Short Runs}
\begin{figure}
\resizebox{\hsize}{!}{\includegraphics{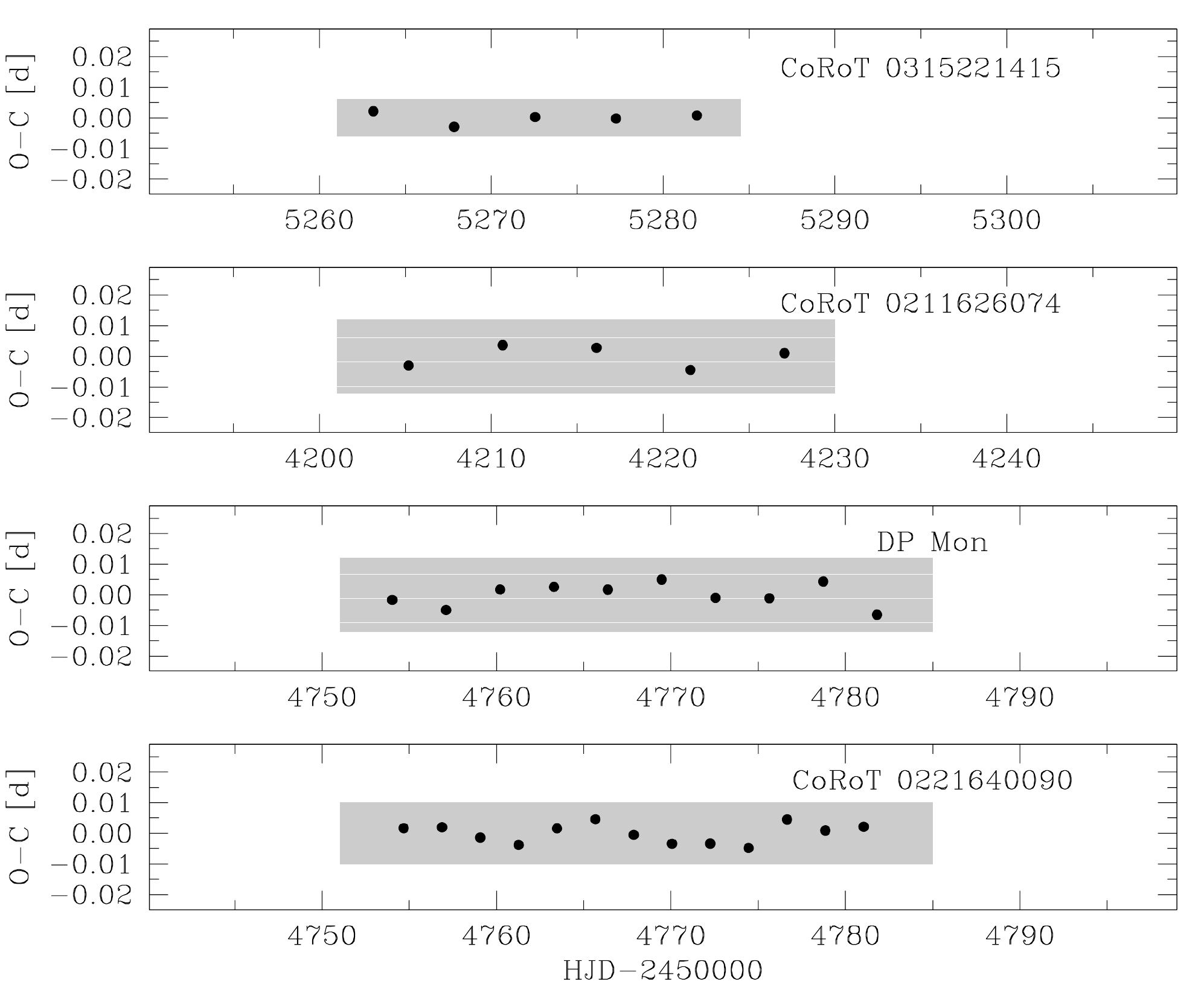}}
\caption{The O-C values with respect to a linear ephemeris fitting the \tmax\, of CoRoT~0221640090,
DP Mon, CoRoT~0315221415, and CoRoT~0211626074.
The grey box is  the $\pm3\sigma$ band (where $\sigma$ is the standard deviation of the fit) 
around the average values.
} 
\label{ocallcep}
\end{figure}

The solutions of the light curves of the Cepheids 
CoRoT~0221640090 (Table~\ref{soliii}), CoRoT~0211626074 (Table~\ref{soliv},
CoRoT~0315221415 (Table~\ref{solv}), and DP Mon (Table~\ref{solvi}
were much less laborious since these stars were observed once in some Short Runs (SRs) and
hence followed for a restricted number of cycles. The O-C plots do not provide useful
hints about periodic variations (Fig.~\ref{ocallcep}). Only in the case of CoRoT~0221640090 
(bottom panel) the  string of consecutive \tmax\, shows a hint of regular behaviour, but the monitoring
is  too short to assess its repetitivity on a long time-scale. Moreover,
the standard deviation around the linear ephemeris is $\pm$0.0033~d and all the maxima are well
within 3$\sigma$.

\begin{table*}
\caption{Fourier parameters of the curves of the Cepheids observed  with CoRoT. See Table~\ref{colori} for \cep.}
\label{others}
\begin{tabular} {ll  lll  llll }
\hline
\multicolumn{1}{c}{Star}&\multicolumn{1}{c}{Mode} &\multicolumn{1}{c}{\tmax [HJD-} &\multicolumn{1}{c}{Period}&\multicolumn{1}{c}{Ampl.}&
\multicolumn{1}{c}{\fdu}&\multicolumn{1}{c}{\rdu}&\multicolumn{1}{c}{\ftu}&\multicolumn{1}{c}{\fqu} \\
\multicolumn{1}{c}{}&\multicolumn{1}{c}{}&\multicolumn{1}{c}{2450000]} &\multicolumn{1}{c}{[d]}&\multicolumn{1}{c}{[mag]}&
\multicolumn{1}{c}{[rad]}&\multicolumn{1}{c}{}&\multicolumn{1}{c}{[rad]}&\multicolumn{1}{c}{[rad]} \\
\hline
\noalign{\smallskip}
\cepdue & $F$            &    4785.4731 & 1.651969(1) & 0.372 & 4.1811(3) & 0.3745(1) & 2.2206 (1) &  0.172(1) \\
CoRoT~0221640090  & $1O$ &    4754.6755 &  2.19758(4) & 0.288 & 4.560(1) &  0.1397(1) &  3.641(5) & 2.670(9) \\
DP Mon & $F$  &               4754.0407 &  3.08704(4)  & 0.678 & 4.312(1) &  0.5201(1) &  2.4703(4)& 0.5439(6) \\
CoRoT~0315221415 & $F$  &     5258.4566  &  4.7085(17) & 0.557 & 4.363(1) &  0.4650(5) &  2.478(3) & 0.434(5) \\
CoRoT~0211626074 & $F$  &     4199.7103  &  5.4706(2)  & 0.296 & 4.731(1) &  0.3147(3) &  3.242(3) & 1.391(35) \\
\noalign{\smallskip}
CoRoT~0659466327 & Pop.~II&   6031.4482  &  1.542353(2)  & 0.492 & 4.394(1) &  0.2913(2) &  0.886(1) & 4.953(2) \\
\hline
\end{tabular}
\end{table*}

\begin{figure}
\resizebox{\hsize}{!}{\includegraphics{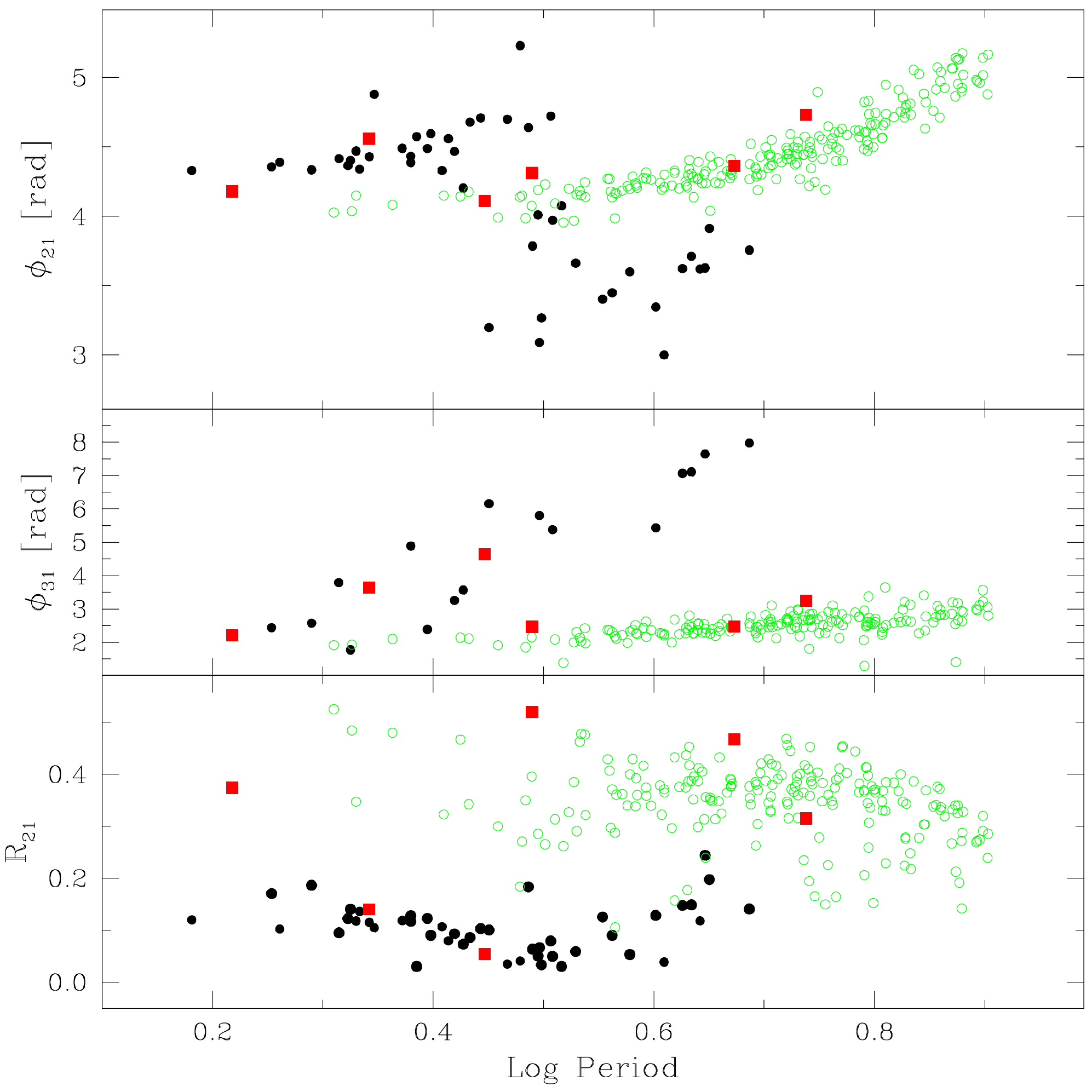}}
\caption{The separation between fundamental (open green circles) and first overtone (filled black circles)
Galactic Pop.~I Cepheids in the progression of the Fourier parameters (in $V$ light) in function of the period. 
The red squares indicate the positions of the CoRoT Cepheids: \cepdue, CoRoT~0221640090, \cep, DP Mon, 
CoRoT~0315221415, and CoRoT~0211626074 (from left to right). 
} 
\label{grafo}
\end{figure}

\section{The Fourier parameters as pulsation mode discriminant}
The Fourier decomposition of the light curves of Cepheids allowed us
to disentangle $F$ and $1O$ radial pulsators (Fig.~\ref{grafo}). 
The \fdu-sequence formed by the classical Cepheids
reproduces  the well-known Hertzsprung
progression. It can be described by the linear relation 
$\phi_{21}=3.332~+~0.216~P$ \citep{cesme}. 
The progression is sharp, since a shift of $\pm$0.30~rad pushes a star well
outside it. Another Z-shaped sequence was
put in evidence by applying the Fourier decomposition to 
the light curves of Galactic Cepheids with $P<$8~d: 
the upper part of this sequence is composed of stars with
\fdu$>$4.2~rad, the lower part of stars with 
\fdu$<$~4.0~rad (Fig.~\ref{grafo}, top panel).
The \ftu$-P$ plane was the Rosetta
stone to link  the upper and lower sequences, since it shows just one continuous sequence,
well separated from the classical one (Fig.~\ref{grafo}, middle panel). The Fourier
decomposition of the light curves of double-mode ($F/1O$) Cepheids provided the definitive
evidence that the pulsation mode drives the separation between the different
sequences \citep{pardo,ccterm}.  The \rdu$-P$ plot shows that the amplitude of the harmonic 2\fu\, 
of the $1O$-mode pulsators is  much smaller than that of the main frequency \fu\,
(Fig.~\ref{grafo}, bottom panel). This fact makes the light curves of the $1O$-mode pulsators
much more (but not exactly) sine-shaped than those of $F$-mode pulsators.
Theoretical works demonstrated how the inclusion of convective energy transport has made possible
to reproduce the observed sequence of $1O$-pulsators and that the sharp features in the
Fourier coefficients are due to the resonance $P_1/P_4=2$ between the fourth and first overtones \citep{feucht}.

The classification of CoRoT~0211626074 and CoRoT~0315221415 is straightforward, since
the \fdu, \rdu, and \ftu\,  parameters put both these long-period (in our sample) Cepheids on the
well-defined sequence of $F$-mode pulsators.

The period of \cep\, and DP Mon are very particular in this context. They lie exactly on the discontinuity
of the Z-shaped sequence of the $1O$ pulsators in the \fdu$-P$ plane, where the  intersection with
the $F$-sequence  occurs. Therefore, the \fdu\, parameter alone is not able to identify
their pulsation mode. However, the much higher value of \ftu\, and the much smaller value
of \rdu\, with respect to those of the classical Cepheids clearly demonstrate that \cep\, is a $1O$ pulsator. 
On the other hand, the \ftu\, and \rdu\, parameters of DP~Mon match the $F$-sequence.
The light curve of DP Mon is very asymmetrical (Fig.~\ref{curve})  and it is very similar to those
of RRab stars. In particular, notice the kink before the minimum, that requires a large
number of harmonics to be fitted.

We  classify CoRoT~0221640090 as a $1O$ pulsator, like \cep. The case
of \cepdue\, is more intriguing, since its period is very short also with respect to the
full sample of Galactic Cepheids. The Cepheids with $P<$2.5~d are quite rare and 
the space of Fourier parameters is scarcely populated in this region.
For this reason the values of \fdu\, and \ftu\, are
not able to provide a clear mode identification. However, the \rdu\, value is quite high, and hence
the star is very probably a $F-$mode pulsator.

\subsection{The coloured light curves}
The continuous photometry allowed us to verify the phases of maximum brightness
($\phi_{\rm max}$) in the CoRoT colours with great accuracy. 
The analyses described in the previous
sections were repeated for each colour of each star. The results are summarized in
Table~\ref{coloriall}. There is a common trend indicating that the maximum brightness
occurs a little earlier at short wavelengths than at long ones. The only exception
is DP~Mon, whose maxima are measured at the same phases (within error bars) in all 
colours. 

The observed phase shifts are very small, around 0.01~period. This result strongly corroborates
 the practice  to merge 
all the maxima observed with different filters or techniques when studying secular
period variations. Indeed, this shift is usually smaller than the uncertainties on the
\tmax\, determination in case of ground-based, fragmented observations. 
This is the first measurement of the phase shifts among light curves of Cepheids in different colours
after the pioneering  approach by \citet{wisn}.

\begin{table}
\caption{Phase shifts of the light curves of the Cepheids observed  in the CoRoT chromatic mode.}
\label{coloriall}
\begin{tabular} {l  rrrr }
\hline
\multicolumn{1}{c}{Star}&\multicolumn{1}{c}{Blue} &\multicolumn{1}{c}{Green} &\multicolumn{1}{c}{Red}&\multicolumn{1}{c}{White}\\
\hline
\cep        &      0.993 &      0.997 &      1.004 &      1.000 \\
in LRa01                 & $\pm$0.003 & $\pm$0.003 & $\pm$0.003 & $\pm$0.003 \\
\noalign{\smallskip}
\cep        &      0.994 &      0.996 &      1.003 &      1.000 \\
in LRa06                 & $\pm$0.005 & $\pm$0.005 & $\pm$0.003 & $\pm$0.003 \\
\noalign{\smallskip}
CoRoT~0221640090 &      0.994 &      1.002 &      1.007 &      1.000 \\
                 &  $\pm$0.003 & $\pm$0.003 & $\pm$0.005 & $\pm$0.003 \\
\noalign{\smallskip}
V376 Mon         &      0.996 &      0.998 &      1.002 &      1.000\\
                 & $\pm$0.003 & $\pm$0.003 & $\pm$0.003 & $\pm$0.003 \\
\noalign{\smallskip}
DP Mon           &      0.998 &      1.000 &      1.001 &      1.000 \\
                 & $\pm$0.003 & $\pm$0.003 & $\pm$0.003 & $\pm$0.003 \\
\hline
\end{tabular}
\end{table}

\section{Discussion and conclusions}

 The power spectra of the CoRoT light curves of Galactic Cepheids did not show any convincing evidence
of the excitation of nonradial modes, especially in the three monitored in the Long Runs.
Stars showing  additional modes were found with an incidence of 10\% among the 
$1O$-Cepheids of the LMC \citep{lmcogle}. The only two $1O$-pulsators observed by CoRoT
cannot yet tell us if
Galactic $1O$-Cepheids behave differently than LMC ones.
On the other hand, the
$F$-Cepheids of the LMC do not show additional modes \citep{lmcogle,mosk} and the
intensive CoRoT photometry does not modify this issue. 
The detection of the same peak in 
two independent sets of residuals of \cep\,  seemed to be 
a decisive piece of evidence in favour of the excitation of a  nonradial mode, but
it not survived the change of the procedure for the computation of the residuals.
 
In our opinion the instrumental effects seem to play an
important role in the long time series of space photometry. 
Our recommendation is to analyse residuals calculated from local solutions
(e.g. cycle-to-cycle when possible) to search for additional periodicities.
Instrumental drifts and jumps (Fig.~\ref{lruns}) or, in extreme cases, numerical errors or amplitude
changes can enhance spurious peaks in the power spectrum of the residuals
calculated from a global fit (Fig.~\ref{spectrares}). The residuals from the cycle-by-cycle
fits are providing a much more reliable dataset to be analysed in frequency (Fig.~\ref{curveres}). 
Omitting a high harmonic from the fit and verify that it is correctly recovered
is an excellent litmus test of the whole procedure.
Moreover,  the lack of combination terms
between the main pulsation mode and the suspected additional component  should be
considered a warning against the identification of this component as a true 
excited  mode.

Another compelling result expected from the analysis of the CoRoT light
curves was to verify the repetitivity and the amplitude of the 
cycle-to-cycle variations, as those observed in the {\it Kepler} 
data of V1154 Cyg \citep{poster} and in the radial-velocity surveys
\citep{anderson}. The Fourier parameters of the light curves of CoRoT
Cepheids show some variability, but in general all the changes are within $\pm3\sigma$.
Moreover, they seem to be erratic, not reproducing phenomena like
activity cycle or long-term effects.

No large scatter was observed around the maxima of the folded light curves and
no large variations were observed in the O-C plots of the Cepheids monitored with 
CoRoT. In particular, the O-C values are all within $\pm$0.01~d, i.e.,$\pm$15~min,
including the $1O$-mode pulsators \cep\, and  CoRoT~0221640090
(Figs.~\ref{fourier}, \ref{decocef751}, \ref{decocef327}, and \ref{ocallcep}).
This is particularly interesting since
the space photometry obtained with the satellite {\it MOST} showed some instabilities in the pulsation
cycle of the $1O$-Cepheid SZ Tau, while the $F$-mode RT Aur repeated the light curve
more precisely \citep{most}. Therefore, we performed some additional tests on
the folded light curves, evaluating 
the standard deviations as a function of phase bins. 
Tiny increases of the standard deviations were measured in correspondence of
the most rapid light variation (i.e., on the steep ascending branch); they 
mostly reflect the variability due to the pulsation within the phase bins, as
proven by the fact that they appear to be more relevant for the stars showing
the largest amplitudes and the most asymmetrical light curves. 
This suggests that the standard deviations of the phase bins on the rising branch cannot be 
used as a tool to search for cycle-to-cycle variations.  
We can conclude that in the case of the seven CoRoT Cepheids, the cycle-to-cycle variability is very limited
and not greatly affecting the repetitivity of the light curves. 
The light variations observed 
around the maxima of SZ Tau are more promising, but on this context the low-frequency peaks in the power
spectrum have to be carefully considered: the residuals observed in {\it MOST} photometry \citep[Fig.~6 in ][]
{most} are very similar to the spurious ones detected in CoRoT one (Fig.~\ref{curveres}).

We also benefit from the unique opportunity provided by the CoRoT multicolour photometry.
Very small shifts in the \tmax 's were observed for the first time in the optical range.
Because these shifts are  around 0.01~period only, our analysis support the use of 
\tmax 's observed in different passbands 
to reconstruct secular period variations \citep[e.g. ][]{most}. 

The accurate values of the Fourier parameters provided by the available space photometry
up to very high harmonics 
can be used to improve the automatic classification of the Cepheid subclasses,
i.e., Pop.~I, Pop.~II,  $F$-mode, and $1O$-mode pulsators. The quantitative separation of these
subclasses in the parameter space is of particular relevance when considering the large harvest of new variables
that will be discovered by future, wide-field, and all-sky space missions like 
TESS \citep{tess} and PLATO~2.0 \citep{plato}.

\section*{Acknowledgements}
The CoRoT space mission has been developed and operated by CNES, with contributions
from Austria, Belgium, Brazil, ESA (RSSD and Science Program), Germany, and Spain.
The authors thank the anonymous referee for useful comments.
EP acknowledges Observatoire Midi-Pyr\'en\'ees for the two-months grant allocated between
2014 May and July; useful discussions with Boris Dintrans and Pascal Fouqu\'e at the
very beginning of the project are acknowledged.
JMB thanks for the support of the NKFIH Grant K-115709.
MR acknowledges financial support from the FP7 project 
{\it SpaceInn: Exploitation of Space Data for Innovative Helio- and Asteroseismology},
EP acknowledges PRIN-INAF 2014 {\it Galactic Archaelogy} for support to latest activities.
This research has made use of the ExoDat Database, operated at LAM-OAMP,
 Marseille, France, on behalf of the CoRoT/Exoplanet program. 
The present study has used the SIMBAD data base operated at the Centre
de Donn\'ees Astronomiques (Strasbourg, France).

\appendix
\section{Solutions of the CoRoT light curves}
\label{appen}
We provide the solutions of the CoRoT light curves of the Cepheids observed in the 
Long Runs. 
We fitted the CoRoT magnitudes by means of the formula
\begin{equation}
m(t)= m_o + \sum_i {A_i \cos [2\pi~i~f  (t-T_o) +\phi_i ]}
\end{equation}
and we calculated
the Fourier parameters $R_{ij}=A_i/A_j$ (e.g. $R_{21}=A_2/A_1$)
and $\phi_{ij}=j~\phi_{i}-i~\phi_{j}$ (e.g. $\phi_{21}=\phi_2 -
2~\phi_1$ and $\phi_{31}=\phi_3 -3~\phi_1$) from the amplitude ($A_i$)
and phase ($\phi_i$) coefficients.
The error bars were calculated at the $1\sigma$ level \citep{dsn}.
\begin{table*}
\caption{Coefficients and formal errors (1$\sigma$) of the least--squares solutions of
the light curves  of the Pop.~I, $1O$-mode pulsator \cep\, and of the 
Pop.~I, $F$-mode pulsator \cepdue.}
\begin{tabular} {r r ll c ll c ll }
\hline
\multicolumn{2}{l}{}& \multicolumn{5}{c}{\cep, \fu=0.3573905(1)~\cd} && \multicolumn{2}{c}{\cepdue} \\
\cline{3-7}
\multicolumn{2}{l}{}&\multicolumn{2}{c}{LRa01} &&\multicolumn{2}{c}{LRa06}&& \multicolumn{2}{c}{\fu=0.6053382(4)~\cd}\\
\cline{3-4} \cline{6-7} \cline{9-10} 
\multicolumn{1}{c}{Term}  && \multicolumn{1}{c}{Ampl.} &
\multicolumn{1}{c}{Phase} && \multicolumn{1}{c}{Ampl.} & \multicolumn{1}{c}{Phase} &&
\multicolumn{1}{c}{Ampl.} & \multicolumn{1}{c}{Phase}\\
\multicolumn{1}{c}{} &   &\multicolumn{1}{c}{[mag]} &
\multicolumn{1}{c}{[rad]} && \multicolumn{1}{c}{[mag]} & \multicolumn{1}{c}{[rad]}&&
\multicolumn{1}{c}{[mag]} & \multicolumn{1}{c}{[rad]} \\
\hline
\fu  &&0.110500(4) & 3.06375(4)     && 0.116054(6) & 3.05962(5)    && 0.15852(2) & 2.5446(1)   \\ 
2\fu &&0.005962(4) & 3.95440(72)    && 0.006552(6) & 3.94414(90)   && 0.05937(2) & 2.9871(2)   \\
3\fu &&0.000699(4) & 1.26965(616)   && 0.000896(6) & 1.30168(655)  && 0.02715(2) & 3.5712(5)   \\
4\fu &&0.000621(4) & 3.31223(693)   && 0.000809(6) & 3.24492(726)  && 0.01096(2) & 4.0676(14)  \\
5\fu &&0.000287(4) & 4.55196(1499)  && 0.000389(6) & 4.51949(1509) && 0.00435(2) & 4.4503(34)  \\
6\fu &&0.000135(4) & 5.74428(3180)  && 0.000120(6) & 5.64753(4876) && 0.00156(2) & 4.6689(94)  \\
7\fu &&0.000049(4) & 0.55276(8822)  && 0.000078(6) & 0.09852(7486) && 0.00077(2) & 4.5913(190) \\
8\fu &&0.000024(4) & 2.10979(17710) && 0.000025(6) & 1.97647(23114)&& 0.00049(2) & 4.6907(297) \\
9\fu     &&            &               &&             &            && 0.00036(2) & 4.9298(408) \\
10\fu     &&            &               &&             &           && 0.00028(2) & 5.3570(534) \\
11\fu     &&            &               &&             &           && 0.00016(2) & 5.8615(952) \\
12\fu     &&            &               &&             &           && 0.00010(2) & 6.1633(1511)\\
13\fu     &&            &               &&             &           && 0.00007(2) & 0.7590(2155)\\
14\fu     &&            &               &&             &           && 0.00005(2) & 0.7422(3075)\\
\hline
\multicolumn{2}{l}{Res. rms}&\multicolumn{2}{c}{0.00174 mag} &&\multicolumn{2}{c}{0.00184 mag}&&
\multicolumn{2}{c}{0.00133 mag} \\
\multicolumn{2}{l}{T$_0$}&\multicolumn{2}{c}{HJD~2454394.6990} && \multicolumn{2}{c}{HJD~2455936.4282}&&
\multicolumn{2}{c}{HJD~2454785.4731}\\
\hline
\label{sol}
\end{tabular}
\end{table*}

\begin{table}
\caption{Coefficients  and formal errors (1$\sigma$) of the least--squares solution of
the light curve  of the Pop.~II Cepheid CoRoT~0659466327.}
\begin{tabular} {r r ll}
\hline
\multicolumn{2}{l}{}& \multicolumn{2}{c}{CoRoT 0659466327}\\
\multicolumn{2}{l}{}& \multicolumn{2}{c}{\fu=0.648366006 ~\cd}\\
\cline{3-4}
\multicolumn{1}{c}{Term}  && \multicolumn{1}{c}{Ampl.} & \multicolumn{1}{c}{Phase}\\
&&\multicolumn{1}{c}{[mag]} & \multicolumn{1}{c}{[rad]}\\
\hline
\fu   && 0.19348(4) & 2.6323(2)\\
2\fu  && 0.05637(4) & 3.3752(7)\\
3\fu  && 0.03478(4) & 2.5001(12)\\
4\fu  && 0.02705(4) & 2.9155(16)\\
5\fu  && 0.01608(4) & 3.8907(26)\\
6\fu  && 0.01154(4) & 4.7951(36)\\
7\fu  && 0.00629(4) & 5.6173(67)\\
8\fu  && 0.00262(4) & 0.3086(161)\\
9\fu  && 0.00055(4) & 1.1627(767)\\
10\fu && 0.00052(4) & 6.0158(832)\\
11\fu && 0.00076(4) & 0.8167(555)\\
12\fu && 0.00083(4) & 1.9510(508)\\
13\fu && 0.00068(4) & 2.9567(614)\\
14\fu && 0.00041(4) & 4.1419(1024)\\
15\fu && 0.00019(4) & 5.2084(2196)\\
16\fu && 0.00007(4) & 5.7712(6216)\\
17\fu && 0.00009(4) & 5.4264(4543)\\
18\fu && 0.00016(4) & 0.2652(2712)\\
\hline
\multicolumn{2}{l}{Res. rms}& \multicolumn{2}{c}{0.00323 mag}\\
\multicolumn{2}{l}{T$_0$}&\multicolumn{2}{c}{HJD~2456031.4482}\\
\hline
\label{solii}
\end{tabular}
\end{table}

\begin{table}
\caption{Coefficients  and formal errors (1$\sigma$) of the least--squares solution of
the light curve  of the Pop.~I Cepheid, $1O$-mode pulsator CoRoT~0221640090.}
\begin{tabular} {r r ll}
\hline
\multicolumn{2}{l}{}& \multicolumn{2}{c}{CoRoT 0221640090}\\
\multicolumn{2}{l}{}& \multicolumn{2}{c}{\fu=0.455047~\cd}\\
\cline{3-4}
\multicolumn{1}{c}{Term}  && \multicolumn{1}{c}{Ampl.} & \multicolumn{1}{c}{Phase}\\
&&\multicolumn{1}{c}{[mag]} & \multicolumn{1}{c}{[rad]}\\
\hline
\fu   && 0.14079(1) & 2.9111(1)\\
2\fu  && 0.01967(1) & 4.0987(7)\\
3\fu  && 0.00290(1) & 6.0910(47)\\
4\fu  && 0.00156(1) & 1.7476(87)\\
5\fu  && 0.00102(1) & 3.1642(132)\\
6\fu  && 0.00055(1) & 4.3818(244)\\
7\fu  && 0.00035(1) & 5.5170(384)\\
8\fu  && 0.00017(1) & 0.4274(795)\\
9\fu  && 0.00009(1) & 1.3975(1468)\\
10\fu && 0.00002(1) & 1.9638(6112)\\
11\fu && 0.00006(1) & 3.2910(2157)\\
12\fu && 0.00002(1) & 3.6209(6119)\\
\hline
\multicolumn{2}{l}{Res. rms}& \multicolumn{2}{c}{0.00259 mag}\\
\multicolumn{2}{l}{T$_0$}&\multicolumn{2}{c}{HJD~2454754.6755}\\
\hline
\label{soliii}
\end{tabular}
\end{table}

\begin{table}
\caption{Coefficients  and formal errors (1$\sigma$) of the least--squares solution of
the light curve  of the Pop.~I Cepheid, $F$-mode pulsator CoRoT~0211626074.}
\begin{tabular} {r r ll}
\hline
\multicolumn{2}{l}{}& \multicolumn{2}{c}{CoRoT 0211626074}\\
\multicolumn{2}{l}{}& \multicolumn{2}{c}{\fu=0.182796~\cd}\\
\cline{3-4}
\multicolumn{1}{c}{Term}  && \multicolumn{1}{c}{Ampl.} & \multicolumn{1}{c}{Phase}\\
&&\multicolumn{1}{c}{[mag]} & \multicolumn{1}{c}{[rad]}\\
\hline
\fu   && 0.12984(3) & 2.5207(3)\\
2\fu  && 0.04086(3) & 3.4897(8)\\
3\fu  && 0.01028(3) & 4.5213(32)\\
4\fu  && 0.00095(3) & 5.1938(341)\\
5\fu  && 0.00166(3) & 3.0263(195)\\
6\fu  && 0.00128(3) & 3.7305(253)\\
7\fu  && 0.00103(3) & 4.5825(315)\\
8\fu  && 0.00062(3) & 5.3574(527)\\
9\fu  && 0.00036(3) & 6.2663(912)\\
10\fu && 0.00018(3) & 0.8783(1842)\\
11\fu && 0.00010(3) & 2.1266(3298)\\
12\fu && 0.00008(3) & 2.7998(4124)\\
\hline
\multicolumn{2}{l}{Res. rms}& \multicolumn{2}{c}{0.00144 mag}\\
\multicolumn{2}{l}{T$_0$}&\multicolumn{2}{c}{HJD~2454199.7103}\\
\hline
\label{soliv}
\end{tabular}
\end{table}

\begin{table}
\caption{Coefficients  and formal errors (1$\sigma$) of the least--squares solution of
the light curve  of the Pop.~I Cepheid, $F$-mode pulsator CoRoT~0315221415.}
\begin{tabular} {r r ll}
\hline
\multicolumn{2}{l}{}& \multicolumn{2}{c}{CoRoT 0315221415}\\
\multicolumn{2}{l}{}& \multicolumn{2}{c}{\fu=0.212384~\cd}\\
\cline{3-4}
\multicolumn{1}{c}{Term}  && \multicolumn{1}{c}{Ampl.} & \multicolumn{1}{c}{Phase}\\
&&\multicolumn{1}{c}{[mag]} & \multicolumn{1}{c}{[rad]}\\
\hline
\fu   && 0.21116(6) & 2.4327(3)\\
2\fu  && 0.09819(6) & 2.9456(6)\\
3\fu  && 0.04415(6) & 3.4927(14)\\
4\fu  && 0.02011(6) & 3.8819(31)\\
5\fu  && 0.00735(6) & 4.0916(84)\\
6\fu  && 0.00386(6) & 3.9842(159)\\
7\fu  && 0.00298(6) & 3.7086(207)\\
8\fu  && 0.00314(6) & 3.8961(196)\\
9\fu  && 0.00299(6) & 4.3491(205)\\
10\fu && 0.00246(6) & 4.8484(248)\\
11\fu && 0.00201(6) & 5.4295(305)\\
12\fu && 0.00158(6) & 6.0745(387)\\
13\fu && 0.00122(6) & 0.3769(501) \\
14\fu && 0.00087(6) & 1.1612(706) \\
15\fu && 0.00060(6) & 1.8127(1016)\\
16\fu && 0.00045(6) & 2.6720(1363)\\
17\fu && 0.00032(6) & 3.5748(1915)\\
18\fu && 0.00016(6) & 4.2500(3960)\\
19\fu && 0.00020(6) & 5.1683(3062)\\
20\fu && 0.00016(6) & 0.3755(3767)\\
\hline
\multicolumn{2}{l}{Res. rms}& \multicolumn{2}{c}{0.00262 mag}\\
\multicolumn{2}{l}{T$_0$}&\multicolumn{2}{c}{HJD~2455258.4566}\\
\hline
\label{solv}
\end{tabular}
\end{table}

\begin{table}
\caption{Coefficients  and formal errors (1$\sigma$) of the least--squares solution of
the light curve  of the Pop.~I Cepheid, $F$-mode pulsator DP~Mon.}
\begin{tabular} {r r ll}
\hline
\multicolumn{2}{l}{}& \multicolumn{2}{c}{DP Mon}\\
\multicolumn{2}{l}{}& \multicolumn{2}{c}{\fu=0.323939~\cd}\\
\cline{3-4}
\multicolumn{1}{c}{Term}  && \multicolumn{1}{c}{Ampl.} & \multicolumn{1}{c}{Phase}\\
&&\multicolumn{1}{c}{[mag]} & \multicolumn{1}{c}{[rad]}\\
\hline
\fu   && 0.23679(2) & 2.2927(1)\\
2\fu  && 0.12315(2) & 2.6145(1)\\
3\fu  && 0.07198(2) & 3.0652(2)\\
4\fu  && 0.04044(2) & 3.4315(4)\\
5\fu  && 0.02352(2) & 3.6996(7)\\
6\fu  && 0.01444(2) & 4.0036(12)\\
7\fu  && 0.00856(2) & 4.2711(20)\\
8\fu  && 0.00523(2) & 4.4842(33)\\
9\fu  && 0.00323(2) & 4.6872(54)\\
10\fu && 0.00205(2) & 4.7897(85)\\
11\fu && 0.00126(2) & 4.8674(137)\\
12\fu && 0.00096(2) & 5.0046(180)\\
13\fu && 0.00074(2) & 5.0513(234) \\
14\fu && 0.00058(2) & 5.2112(296) \\
15\fu && 0.00048(2) & 5.3839(359)\\
16\fu && 0.00041(2) & 5.6321(425)\\
17\fu && 0.00034(2) & 5.9751(515)\\
18\fu && 0.00024(2) & 6.0823(712)\\
19\fu && 0.00022(2) & 0.1627(799)\\
20\fu && 0.00014(2) & 0.5256(1234)\\
21\fu && 0.00016(2) & 0.7288(1113)\\
22\fu && 0.00010(2) & 1.3601(1793)\\
23\fu && 0.00009(2) & 1.1871(2008)\\
24\fu && 0.00003(2) & 1.4777(5286)\\
25\fu && 0.00005(2) & 1.8188(3629)\\
\hline
\multicolumn{2}{l}{Res. rms}& \multicolumn{2}{c}{0.00328 mag}\\
\multicolumn{2}{l}{T$_0$}&\multicolumn{2}{c}{HJD~2454754.0407}\\
\hline
\label{solvi}
\end{tabular}
\end{table}
\label{lastpage}

\begin{thebibliography}{mnras}
\bibitem[Anderson(2014)]{anderson} Anderson R.I., 2014, \aap, 566, L10
\bibitem[Antonello, Poretti \& Reduzzi(1990)]{apr} Antonello E., Poretti E.,  Reduzzi L., 1990, \aap, 236, 138 
\bibitem[Benk\H{o} et al.(2014)]{benko} Benk\H{o} J.M., Plachy E., Szab\'o R., Moln\'ar L., Koll\'ath Z.,
 2014, \apjs,  213, 31
\bibitem[Berdnikov et al.(2014)]{berdni} Berdnikov L.N., Kniazev A.Yu., Sefako R., Kravtsov V.V.,  Zhujko S.V.,
2014, Astronomy Letter, 40, 125
\bibitem[Bruntt et al.(2008)]{bruntt} Bruntt H., et al., 2008, \apj, 683, 433
\bibitem[Buchler(2008)]{buchler} Buchler J.R., 2008, \apj, 680, 1412
\bibitem[Buchler \& Szabo (2007)]{bszabo} Buchler J.R., Szab\'o R., 2007, \apj, 660, 723
\bibitem[Burki, Mayor \& Benz(1982)]{burki} Burki G., Mayor M., Benz W., 1982, \aap, 109, 258
\bibitem[Carpino, Milani \& Nobili(1987)]{mtrap} Carpino M., Milani A., Nobili A.M., 1987, A\&A, 181, 182
\bibitem[Debosscher et al.(2009)]{cvc} Debosscher J.,  et al., 2009, \aap, 506, 519
\bibitem[Deleuil et al.(2009)]{exodat} Deleuil M.,  et al.,  2009, \aj, 138, 649
\bibitem[Derekas et al.(2012)]{v1154} Derekas A., et al., 2012, \mnras, 425, 1312
\bibitem[Donati et al.(1997)]{lsd} Donati J.-F., Semel M., Carter B.D., Rees D.E., Collier Cameron A.,
1997, \mnras, 291, 658
\bibitem[Evans, Sasselov \& Short(2002)]{evans} Evans N.R., Sasselov D.D.,  Short I.S., 2002, \apj, 567, 1121
\bibitem[Evans et al.(2015)]{most} Evans N.R., et al., 2015, \mnras, 446, 4008
\bibitem[Feuchtinger, Buchler \& Koll\`ath(2000)]{feucht} Feuchtinger M., Buchler J.R., Koll\`ath Z., 2000, \apj, 544, 1056
\bibitem[Fruth et al.(2012)]{bestVI} Fruth T.,  et al., 2012, \aj, 143, 140 
\bibitem[Gruberbauer et al.(2007)]{gruber} Gruberbauer M., et al., 2007, \mnras, 379, 1498
\bibitem[Gustafsson et al.(2008)]{marcs} Gustafsson B., Edvardsson B., Eriksson K., J\o rgensen U.G., Nordlund \AA., Plez B., 
2008, \aap, 486, 951
\bibitem[Kabath et al.(2007)]{bestII} Kabath P., Eigm\"uller P., Erikson A., Hedelt P., 
Rauer H., Titz R., Wiese T., Karoff C., 2007, \aj, 134, 1560
\bibitem[Kabath et al.(2008)]{bestIII} Kabath P., et al., 2008, \aj, 136, 654
\bibitem[Kabath et al.(2009a)]{bestIV} Kabath P.,  et al., 2009, \aj, 137, 3911
\bibitem[Kabath et al.(2009b)]{bestV} Kabath P., et al., 2009, \aap, 506, 569
\bibitem[\protect\citeauthoryear{Kanev, Savanov \& Sackhov}{Kanev et al.}2015]{poster} Kanev E., Savanov I., Sachkov M., 2015, in 3$^{\rm rd}$ CoRoT Symposium, 7$^{\rm th}$ KASC Meeting,
The Space Photometry Revolution, Toulouse, July 2014, in press
\bibitem[Karoff et al.(2007)]{bestI} Karoff C.,  et al., 2007, \aj, 134, 766
\bibitem[Kuschnig et al.(1997)]{snr4} Kuschnig R., Weiss W.W., Gruber R., Bely P.Y., Jenkner H., 
1997, \aap, 328, 544
\bibitem[Moln\'ar \&  Szabados(2014)]{molnar} Moln\'ar L.,  Szabados L., 2014, \mnras, 442, 3222
\bibitem[Montgomery \& O'Donoghue(1999)]{dsn} Montgomery M.H., O'Donoghue D., 1999,
Delta Scuti Star Newsletter, 13, 28
\bibitem[Morgan, Simet \& Bargenquast(1998)]{morgan} Morgan S.M., Simet M., Bargenquast, S., 1998,
\actaa, 48, 509
\bibitem[Moskalik(2014)]{pavel} Moskalik P., 2014, in Guzik~J.A., Chaplin~W.J., Handler~G., Pigulski~A., eds., 
IAU Proceedings S301, Vol. 9, Precision Asteroseismology, p.~249
\bibitem[Moskalik \& Ko\l aczkowski(2008)]{lmcogle} Moskalik P., Ko\l aczkowski Z., 2008, Comm. in Asteroseismology, 157, 343
\bibitem[Moskalik \& Ko\l aczkowski(2009)]{mosk} Moskalik P., Ko\l aczkowski Z., 2009, \mnras, 394, 1649
\bibitem[Neilson \& Ignace(2014)]{neilson} Neilson H.R., Ignace R., 2014, \aap, 563, L4
\bibitem[Pardo \& Poretti(1997)]{pardo} Pardo I., Poretti E., 1997, \aap, 324, 121
\bibitem[Poretti(2000)]{cesme} Poretti E., 2000, in Ibanoglu C., ed., 
Nato Science Series, Vol. 544, Variable Stars as Essential Astrophysical Tools, Kluwer Academic Publishers, p.~421
\bibitem[\protect\citeauthoryear{Poretti, Baglin, \& Weiss}{Poretti et al.}2014]{pbw} Poretti E., Baglin A., Weiss W.W., 2014, \apj, 795, L36
\bibitem[Poretti, Le Borgne, \& Klotz(2015)]{kasc} Poretti E., Le Borgne J.F., Klotz A., 2015, in
3$^{\rm rd}$ CoRoT Symposium, 7$^{\rm th}$ KASC Meeting,
The Space Photometry Revolution, Toulouse, July 2014, preprint (arXiv:1506.08122)
\bibitem[Poretti \& Pardo(1997)]{ccterm} Poretti E. \& Pardo I., 1997, \aap, 324, 133
\bibitem[Poretti et al.(2011)]{hads} Poretti E.,  et al., 2011, \aap,  528, A147
\bibitem[Poretti et al.(2005)]{gsc} Poretti E.,  et al., 2005, \aap, 440, 1097 
\bibitem[Rainer(2003)]{laurea} Rainer M., 2003, Laurea Thesis, Universit\`a degli Studi di Milano
\bibitem[Rauer et al.(2014)]{plato} Rauer H., et al., 2014, Exp. Astron., 38, 249
\bibitem[Reinsch(1967)]{reinsch} Reinsch H., 1967, Numerische Mathematik 10, 177
\bibitem[Ricker et al.(2014)]{tess} Ricker G.R.,  et al., 2014, SPIE Proc., 9143, 20
\bibitem[Spreckley \& Stevens(2008)]{smei} Spreckley S.A.,  Stevens I.R., 2008, \mnras, 388, 1239
\bibitem[Szil\'adi  et al.(2007)] {kate} Szil\'adi K., Vink\'o J., Poretti E., Szabados L., Kun M., 2007,
\aap, 473, 579
\bibitem[Valenti \& Fischer(2005)]{valenti} Valenti J.A., Fischer D.A., 2005, \apjs, 159 141 
\bibitem[Valenti \& Piskunov(1996)]{sme} Valenti J.A., Piskunov N., 1996, \aaps, 118, 595
\bibitem[Vani\^cek(1971)]{vani} Vani\^cek P., 1971, \apss, 12, 10
\bibitem[Wachmann(1964)]{arno} Wachmann A.A., 1964, Astron. Abhandlungen Hamburger Sternwarte, 7, 4
\bibitem[Wisniewski \& Johnson(1968)]{wisn} Wisniewski W.Z., Johnson H.L., 1968, 
Communications of the Lunar and Planetary Laboratory, Volume 7, Part 2, p.~57-78
\bibitem[Wu(2001)]{wu} Wu Y., 2001, \mnras, 323, 248
\end{thebibliography}
\end{document}